\documentclass[12pt,draftclsnofoot,onecolumn]{IEEEtran}
\usepackage{multirow}
\usepackage{booktabs}
\usepackage[colorlinks,
            linkcolor=blue,
            anchorcolor=blue,
            citecolor=green
            ]{hyperref}
\usepackage{xcolor}
\usepackage{colortbl}
\usepackage{graphicx}
\usepackage{color}
\usepackage{placeins}
\usepackage{float}
\usepackage{tabularx,colortbl}
\usepackage{amssymb}
\usepackage{amsthm}
\usepackage{cite}
\usepackage{amsmath}
\usepackage[caption=false,font=normalsize,labelfont=sf,textfont=sf]{subfig}
\usepackage{stfloats}
\usepackage{amsthm}
\usepackage{caption2}
\usepackage{bbding}
\usepackage{pifont}
\usepackage{wasysym}
\usepackage{utfsym}
\usepackage{fontawesome}
\usepackage{algorithm}
\usepackage{algpseudocode}
\def\BibTeX{{\rm B\kern-.05em{\sc i\kern-.025em b}\kern-.08em
  T\kern-.1667em\lower.7ex\hbox{E}\kern-.125emX}}
\usepackage{balance}
\begin{document}

\title{Near-field Spatial-domain Channel Extrapolation for XL-MIMO Systems}
\author{Jiayi Lu, Jiayi Zhang, {\it{Senior Member, IEEE}}, Hao Lei, Huahua Xiao, 
\\Bo Ai,~\IEEEmembership{Fellow,~IEEE}, and Derrick Wing Kwan Ng,~\IEEEmembership{Fellow,~IEEE}
\thanks{J. Lu, H. Lei, and B. Ai are with the State Key Laboratory of Advanced Rail Autonomous Operation, and also with the School of Electronics and Information Engineering, Beijing Jiaotong University, Beijing 100044, P. R. China. (e-mail: jiayilu@bjtu.edu.cn).}
\thanks{J. Zhang is with the State Key Laboratory of Advanced Rail Autonomous Operation, Nanjing Rongcai Transportation Technology Research Institute Co., Ltd., Nanjing 210000, China, and also with the School of Electronics and Information Engineering, Beijing Jiaotong University, Beijing 100044, P. R. China. (e-mail: jiayizhang@bjtu.edu.cn).}
\thanks{H. Xiao is with ZTE Corporation and the State Key Laboratory of Mobile Network and Mobile Multimedia Technology. (e-mail: xiao.huahua@zte.com.cn).}
\thanks{D. W. K. Ng is with the School of Electrical Engineering and Telecommunications, University of New South Wales, Sydney, NSW 2052, Australia (e-mail: w.k.ng@unsw.edu.au).}}
\vspace{-6em}
\maketitle
\begin{abstract}
Extremely large-scale multiple-input multiple-output (XL-MIMO) systems are pivotal to next-generation wireless communications, where dynamic RF chain architectures offer enhanced performance. However, efficient precoding in such systems requires accurate channel state information (CSI) obtained with low complexity. To address this challenge, spatial-domain channel extrapolation has attracted growing interest. Existing methods often overlook near-field spherical wavefronts or rely heavily on sparsity priors, leading to performance degradation. In this paper, we propose an adaptive near-field channel extrapolation framework for multi-subcarrier XL-MIMO systems, leveraging a strategically selected subset of antennas. Subsequently, we develop both on-grid and off-grid algorithms, where the latter refines the former's estimates for improved accuracy. To further reduce complexity, a cross-validation (CV)-based scheme is introduced. Additionally, we analytically formulate the mutual coherence of the sensing matrix and propose a coherence-minimizing-based random pattern to ensure robust extrapolation. Numerical results validate that the proposed algorithms significantly outperform existing methods in both extrapolation accuracy and achievable rate, while maintaining low computational complexity. In particular, our proposed CV ratio offers a flexible trade-off between accuracy and efficiency, and the corresponding off-grid algorithm achieves high accuracy with complexity comparable to conventional on-grid methods.
\color{black}
\end{abstract}


\section{Introduction}\label{se:model}
\IEEEPARstart{M}{assive} multiple-input multiple-output (mMIMO) technology has been widely recognized as one of the core technologies for the fifth-generation (5G) mobile communication systems \cite{OBETrans}, \cite{unified_kwan}, delivering substantial beamforming and spatial multiplexing gains to achieve higher transmission rates. As a natural evolution of mMIMO, extremely large-scale MIMO (XL-MIMO) is envisioned to satisfy the stringent demands for low latency, high data rate, and enhanced spectral efficiency in emerging sixth-generation (6G) wireless networks \cite{survey_wz}. Moreover, to integrate hundreds or even thousands of antennas within a compact XL-MIMO array, the millimeter-wave (mmWave) and terahertz (THz) frequency bands are expected to be important operational scenarios for XL-MIMO systems \cite{survey_wz}, \cite{terahertz_dusit}. Furthermore, XL-MIMO distinguishes itself through its excellent capability of achieving pencil-like energy-focused beamforming, effectively mitigating the severe path loss inherent to high-frequency operations \cite{hybrid_zhang}.
\par Notably, XL-MIMO not only scales up the number of antennas but also introduces a paradigm shift in electromagnetic characteristics \cite{fundamentals_wz}, requiring spherical wave modeling to accurately capture near-field effects within the corresponding Rayleigh distance \cite{frauhofer_KT}, \cite{wideband_dai}. Additionally, near-field phenomena enhance the degrees of freedom (DoF) in XL-MIMO channels \cite{survey_wz}, rendering conventional hybrid architectures with a fixed number of radio frequency (RF) chains inadequate for fully leveraging the available channel DoF. To overcome this challenge, a dynamic RF chain architecture was proposed in \cite{precoding_dai}, where multiple RF chains are dynamically configured through a selection network, achieving near-optimal capacity without incurring additional power consumption. Moreover, to unleash the full performance advantages of dynamic XL-MIMO systems, accurate channel state information (CSI) acquisition with low computational complexity is crucial \cite{XL_liu}, \cite{nsn_zhang}. To tackle this challenge, spatial-domain channel extrapolation has gained prominence as a viable approach. Specifically, by exploiting the shared propagation environment and the deterministic arrangement of antennas, this technique enables the reconstruction of the complete array's channel leveraging measurements only obtained from a strategically selected subset of antennas.
\color{black}
\vspace{-1em}
\subsection{Literature Review}
\par Recently, several studies have explored spatial-domain channel extrapolation through artificial intelligence (AI) and compressed sensing (CS) techniques. In the AI-based category, the authors in \cite{mapping_alra} demonstrated that a channel-to-channel mapping function exists when the mapping from candidate user locations to the channels measured by the first antenna subset is bijective. Building upon this insight, deep neural networks (DNNs) were utilized to learn this intricate mapping function given sufficient observations \cite{mapping_alra}, \cite{DNNtask_liu}. Besides, in \cite{8e_gao}, an antenna-domain extrapolation network was proposed, leveraging a constrained degradation algorithm to generate a differentiable approximation for discrete antenna selection. In addition, considering both temporal- and spatial-domain channel extrapolation, the study in \cite{aitfs_zhang} proposed a novel spatial-domain extrapolation method leveraging self-attention and transfer learning. This method utilizes position embedding to preserve the spatial relationships among antennas and employs temporal-domain channel extrapolation data for model pre-training, thereby capturing essential correlations among antenna CSIs. The model is subsequently fine-tuned adopting spatial-domain extrapolation data, with the initial CSI values estimated via linear interpolation.
\par The second category encompasses CS-based methods. For instance, in \cite{multi_han}, a trivariate Newtonized orthogonal matching pursuit (NOMP)-based multi-domain channel extrapolation scheme was proposed to reconstruct channels across various domains, including time, frequency, polarization, spatial, and user dimensions. Furthermore, an enhanced near-field NOMP estimator was developed in \cite{near-reconstruction_han}, extending the methodology proposed in \cite{multi_han} to reconstruct both user equipment (UE)-to-reconfigurable intelligent surface (RIS) and RIS-to-base station (BS) near-field channels. Moreover, the authors in \cite{near-reconstruction_han} also investigated different RIS pattern configurations in terms of radiation profiles and parameter estimation accuracy, demonstrating the superiority of a random configuration strategy. In addition, sparse Bayesian learning (SBL) has been recognized as another powerful approach for channel extrapolation. In particular, a tensor-based SBL framework was proposed in \cite{SBL_zhang} to effectively recover the physical channel parameters and reconstruct the cascaded RIS-assisted MIMO-orthogonal frequency-division multiplexing (OFDM) channels.
\vspace{-1em}
\subsection{Main Contributions}
\par Despite the promising advances discussed above, several critical limitations remain unaddressed in existing works: Firstly, most studies on spatial-domain channel extrapolation rely heavily on learning-based architectures. However, the limited interpretability of these neural networks often hinders a clear understanding of why the selected antenna elements yield optimal extrapolation results and how these elements are spatially related.
Moreover, although extensive research has been conducted on far-field spatial-domain channel extrapolation, spanning a wide range of methods \cite{8e_gao}-\!\!\cite{multi_han}, \cite{SBL_zhang}, \cite{extrapolationinmove_gao}, as well as hardware prototyping and experimental evaluations \cite{HardwareProto_gao}, these approaches are based on the planar wave propagation assumption. Indeed, this assumption becomes inaccurate in near-field scenarios, resulting in noticeable performance degradation. 
This limitation is particularly critical for CS-based methods, where the energy of a single near-field path component spreads over multiple angles \cite{polar_dai}, \cite{holographic_di}, thereby violating the angular-domain sparsity assumption that is fundamental to far-field models.
Last but not least, current CS-based methods, e.g., \cite{wideband_dai}, \cite{multi_han}, and \cite{secure_kwan}, developed heavily rely on prior knowledge of the number of propagation paths, which is often challenging to determine in advance. As highlighted in \cite{adaptive_zhang} and \cite{hierarchical_cai}, employing a fixed number of iterations is generally a suboptimal strategy for achieving both high accuracy and processing efficiency. 
\vspace{-1em}
\begin{table*}[ht]
\caption{Comparison Of Relevant Literature with this Paper.}
\centering
\begin{tabular}{|c|c|c|c|c|c|c|}
\hline
\textbf{References} & \textbf{Near-field} & \textbf{Multi-user} & \textbf{Off-grid} & \textbf{Independent of the} & \textbf{Cross-validation} & \textbf{Deep learning} \\
 &  &  &  & \textbf{number of paths} & \textbf{-based} & \\
\hline
\cite{wideband_dai} & \Checkmark & \XSolidBrush & \XSolidBrush & \XSolidBrush & \XSolidBrush & \XSolidBrush \\
 \hline
\cite{8e_gao} & \XSolidBrush & \Checkmark & - & - & \XSolidBrush & \Checkmark \\
 \hline
\cite{aitfs_zhang} & \Checkmark & \XSolidBrush & - & - & \XSolidBrush & \Checkmark \\
 \hline
\cite{multi_han} & \XSolidBrush & \Checkmark & \Checkmark & \XSolidBrush & \XSolidBrush & \XSolidBrush \\
 \hline
\cite{near-reconstruction_han} & \Checkmark & \XSolidBrush & \Checkmark & \XSolidBrush & \XSolidBrush & \XSolidBrush \\
 \hline
\cite{polar_dai} & \Checkmark & \Checkmark & \Checkmark & \XSolidBrush & \XSolidBrush & \XSolidBrush \\
\hline
Proposed P-ASOMP & \Checkmark & \Checkmark & \XSolidBrush & \Checkmark & \XSolidBrush & \XSolidBrush \\
\hline
Proposed P-ASIGW & \Checkmark & \Checkmark & \Checkmark & \Checkmark & \XSolidBrush & \XSolidBrush \\
\hline
Proposed CV-P-ASOMP & \Checkmark & \Checkmark & \XSolidBrush & \Checkmark & \Checkmark & \XSolidBrush \\
\hline
Proposed CV-P-ASIGW & \Checkmark & \Checkmark & \Checkmark & \Checkmark & \Checkmark & \XSolidBrush \\
\hline
\end{tabular}
\label{tab1}
\end{table*}
\par Motivated by the aforementioned limitations of existing studies, this paper investigates near-field channel extrapolation in a multi-user, multi-subcarrier mmWave XL-MIMO system without relying on prior knowledge of the number of propagation paths. To address the challenges associated with high-dimensional near-field channel reconstruction, we propose a CS-based adaptive channel extrapolation framework. Within this framework, both on-grid and off-grid channel extrapolation algorithms are developed, where the off-grid scheme iteratively refines the coarse on-grid estimates to improve accuracy. Furthermore, a coherence-minimizing antenna selection strategy is proposed to ensure a well-conditioned sensing matrix and enhance extrapolation robustness. The comparisons of relevant literature with this paper are summarized in TABLE~\ref{tab1}. Our main contributions are outlined as follows.

\begin{itemize}
\item A compressed sensing-based adaptive near-field spatial-domain channel extrapolation framework is proposed for mmWave XL-MIMO systems, operating effectively without relying on prior knowledge of the exact path component configuration. Subsequently, in multi-subcarrier scenarios, we formulate the channel extrapolation problem as a sparse signal recovery task under the multiple measurement vector (MMV) scheme. More importantly, we employ a polar-domain dictionary to capture the correlations among antennas, enabling accurate channel reconstruction by jointly estimating both angle and distance information.
\item In the proposed channel extrapolation framework, the sensing matrix is fundamentally determined by the selected antenna elements. Therefore, the antenna selection strategy plays a crucial role in ensuring effective extrapolation performance. To address this, we introduce a novel coherence-minimizing antenna selection strategy, which aims to maintain stable and robust extrapolation performance. Additionally, we evaluate four distinct antenna selection setups, including both deterministic and random antenna positions, by analyzing their radiation profiles and channel extrapolation accuracy. Numerical results are presented to validate the high angular and distance resolution achieved by the proposed strategy.
\item Within the proposed adaptive framework, we introduce an on-grid polar-domain adaptive simultaneous orthogonal matching pursuit (P-ASOMP) algorithm for channel extrapolation. To further enhance extrapolation accuracy, we propose an off-grid polar-domain adaptive simultaneous iterative gridless weighted (P-ASIGW) algorithm that refines near-field channel parameters. To reduce the computational complexity of these adaptive extrapolation algorithms, a cross-validation (CV)-based scheme is developed within the MMV framework. Utilizing this scheme, two enhanced algorithms, CV-P-ASOMP and CV-P-ASIGW, are proposed. By reducing the matrix dimension during the correlation process and incorporating the strong Wolfe conditions, the off-grid CV-P-ASIGW method achieves significantly superior performance, while maintaining comparable computational complexity to existing on-grid methods.
\end{itemize}
\par {\it{Notation:}} Lower-case and upper-case boldface letters ${\mathbf{a}}$ and ${\mathbf{A}}$ denote a vector and a matrix, respectively. The superscripts ${\left( \cdot \right)^\mathrm{T}}$, ${\left( \cdot \right)^\mathrm{H}}$, ${(\cdot)^\dagger}$, and ${\|\cdot\|_\mathrm{F}}$ denote the transpose, conjugate-transpose, pseudo-inverse, and Frobenius norm, respectively. $\odot$ represents the Hadamard product. The $M\times{N}$ real-valued matrix and the $M\times{N}$ complex-valued matrix are denoted by $\mathbb{R}^{M\times{N}}$ and $\mathbb{C}^{M\times{N}}$, respectively. ${\left[\mathbf{A}\right]_{i,j}}$ corresponds to the ${(i,j)}$-th element of ${\mathbf{A}}$. A unit matrix of size $N$ is represented by ${\mathbf{I}_N}$. We denote the circularly symmetric complex-valued Gaussian distribution by ${\mathcal{CN}\left(\mu,\sigma^2\right)}$, where $\mu$ and ${\sigma^{2}}$ denote the mean and the variance. The integer ceiling is denoted as ${\lceil \cdot \rceil}$, and the integer floor is denoted as ${\lfloor \cdot \rfloor}$. The number of combinations for selecting $K$ elements from a set of $N$ elements is denoted as $\mathrm{C}_{N}^{K}$. $\mathbb{E}\{x\}$ is the expectation of a random variable $x$.

\section{System and Channel Model}\label{se:model}
In this paper, we consider an uplink time division duplex (TDD) scenario for an mmWave dynamic XL-MIMO system utilizing OFDM to serve $U$ single-antenna users with $M$ subcarriers\footnote{It is worth mentioning that the multicarrier OFDM framework adopted in this study is readily extendable to orthogonal frequency-division multiple access (OFDMA)-based systems. Specifically, for subcarriers assigned to a single user, the signal model and assumptions made in this paper remain valid.}\color{black}. For the uplink, we assume that the BS receives mutually orthogonal pilot sequences from all $U$ users. This ensures that the channel extrapolation process can be performed independently for each user \cite{polar_dai}. Without loss of generality, we focus our analysis on an arbitrary user. 
\par
\begin{figure*}
\begin{minipage}[t]{0.63\linewidth}
\centering
\includegraphics[width=1\linewidth]{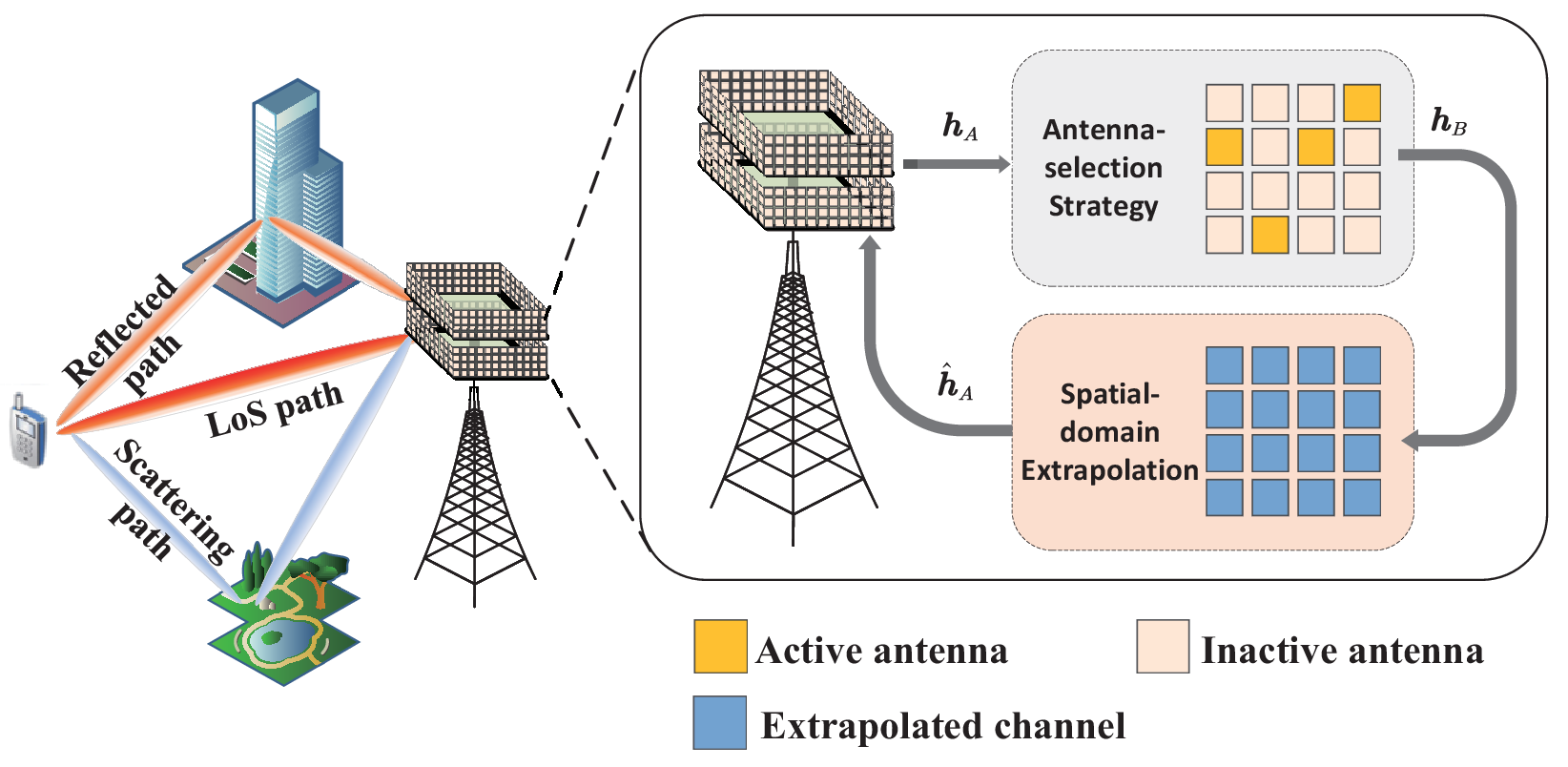}
\caption{\small{Illustration of the spatial-domain channel extrapolation in XL-MIMO systems.}\protect\label{Figure_1}}
\end{minipage}
\begin{minipage}[t]{0.37\linewidth}
\centering
\includegraphics[width=1\linewidth]{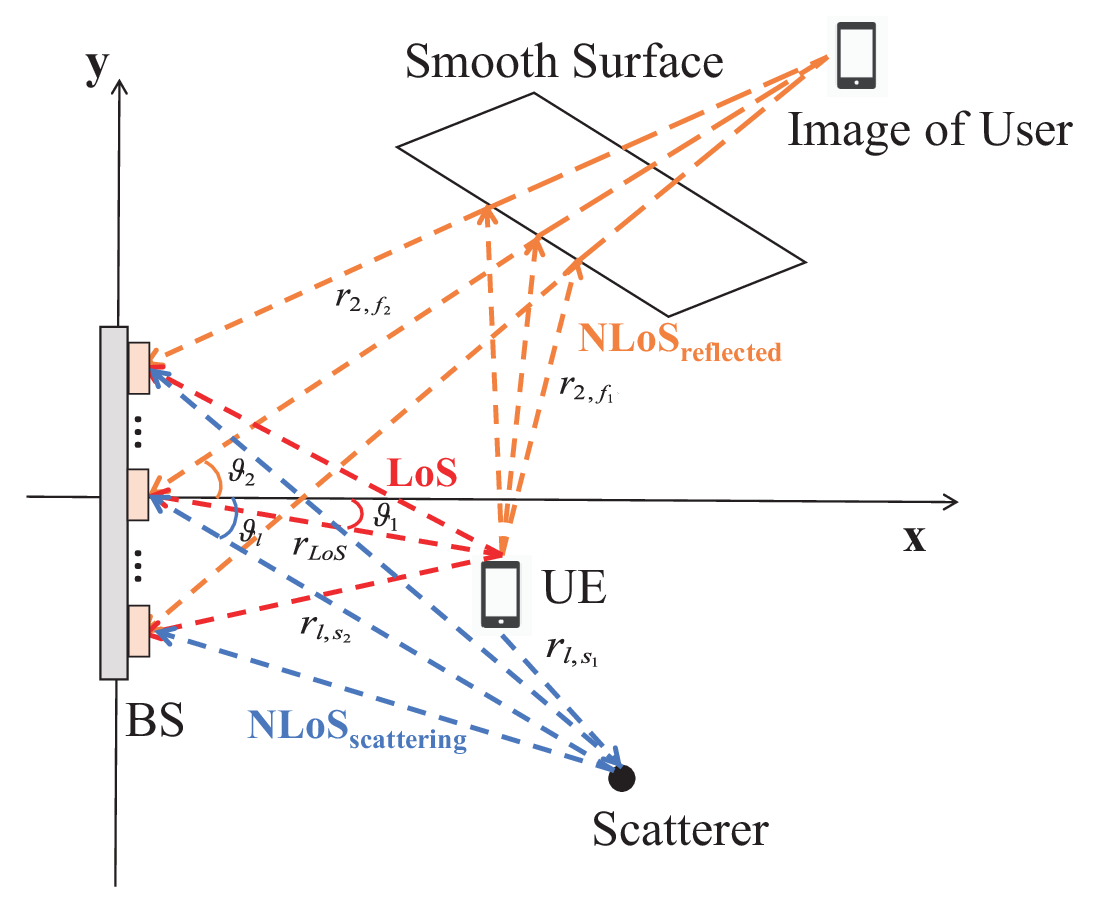}
\caption{\small{Near-field XL-MIMO channel model.}\protect\label{Figure_2}}
\end{minipage}
\end{figure*}

 As illustrated in Fig.~\ref{Figure_1}, the BS employs a uniform linear array (ULA) with a dynamic architecture\footnote{Moreover, it is worth highlighting that the proposed approach is not restricted to the dynamic RF chain configuration described here; rather, it can be readily extended to more general hybrid analog–digital architectures by employing standard on/off switching circuits.}, consisting of $N$ antennas and $K \leq {N}$ RF chains. 
Moreover, as the aperture of the XL-MIMO array increases, the near-field region expands, and the amplitude and phase variations across the array become non-negligible, as shown in Fig.~\ref{Figure_2}. Consequently, the conventional far-field planar wave model becomes inadequate and the near-field spherical wave model should be employed instead. Specifically, the near-field multi-path propagation is modeled as comprising $L$ components, including one line-of-sight (LoS) path and $L\!-\!1$ non-line-of-sight (NLoS) paths \cite{sensing_kwan}. 
\color{black}
Under the spherical wave assumption, the near-field channel vector $\mathbf{h}_m\!\!\in\!\!\mathbb{C}^{N\!\times\!1}$ can be denoted as \cite{channel}
\vspace{-0.5em}\begin{align}
\begin{aligned}
  \mathbf{h}_m&=\sum_{l=1}^{L}g_{l,m}\mathbf{c}(\vartheta_{l},r_{l})\odot\mathbf{b}_m(\vartheta_{l},r_{l})\\
  &=\sum_{l=1}^{L}g_{l,m}\mathbf{a}_m(\theta_{l},r_{l}),
\end{aligned}
\end{align}
where $\theta_{l}=\sin \vartheta_{l}$. In (1), the gain correction vector $\mathbf{c}(\vartheta_{l},r_{l})\in\mathbb{C}^{N\times1}$ can be written as $\mathbf{c}\left(\vartheta_{l},r_{l}\right)=\left[\frac{r_{l}}{r_{l}^{(1)}},\ldots,\frac{r_{l}}{r_{l}^{(N)}}\right]^\mathrm{T}$, where $r_l^{(n)}$ denotes the actual distance from the scatterer to the $n$-th receiving antenna at the BS in the $l$-th path. As shown in Fig.~\ref{Figure_2}, for the LoS path, the gain can be expressed as $g_{1,m}=\frac{\lambda_me^{-jk_{m}{r}_{1}}}{4\pi{r_{1}}}$, where ${k_{m}=\frac{2\pi f_{m}}{c}}$ is the wavenumber, ${\lambda_{m}}$ is the wavelength at frequency $f_m$, and $r_{1}=r_\mathrm{LoS}$ denotes the distance of the LoS path. Besides, the frequency of the $m$-th subcarrier, given a center frequency $f_c$ and system bandwidth $B$, is calculated as ${f_m=f_c+\frac{B}{M}(m-1-\frac{M-1}{2})}$, $m\in\{1,2,\dots,M\}$. 
Moreover, the NLoS paths are composed of a reflected path and $L-2$ scattering paths [27]. The reflected path is typically caused by ground or building surface reflections, as illustrated in Fig.~\ref{Figure_2}. 
\color{black}
The gain of the reflected path is given by $g_{l,m}=\frac{\gamma_{f}\lambda_me^{-jk_{m}({r}_{{l},f_1}+{r}_{{l},f_2})}}{4\pi{({r}_{{l},f_1}+{r}_{{l},f_2})}}$ for $l=2$, where $|\gamma_{f}|\!\in\!(0,1)$ defines the gain degradation due to reflection; ${r}_{{l},f_1}$ and ${r}_{{l},f_2}$ represent the distance from the user to the smooth surface and the distance from the smooth surface to the reference antenna, respectively. The gains of scattering paths are given by $g_{l,m}\!=\!\frac{\gamma_{s}\lambda^2_me^{-jk_{m}({r}_{{l},s_1}+{r}_{{l},s_2})}}{(4\pi)^2{{r}_{{l},s_1} {r}_{{l},s_2}}}$ for $l=3,\dots,L$, where $|\gamma_{s}|\!\in\!(0,1)$ characterizes the reflectivity coefficient of the near-field scatterers; ${r}_{{l},s_1}$ and ${r}_{{l},s_2}$ denote the distance from the user to the scatterer and the distance from the scatterer to the reference antenna, respectively. Therefore, the multipath component gains in the channel model generally satisfy $|g_\mathrm{LoS}| >|g_\mathrm{reflected} | > |g_\mathrm{scattering}|$. 
\color{black}
The near-field array steering vector ${\mathbf{b}_m(\vartheta_{l},r_{l})}\in\mathbb{C}^{N\times1}$ can be represented by \cite{polar_dai}
\vspace{-1em}\begin{align}
\mathbf{b}_m(\vartheta_{l},r_{l})\!=\![e^{\!-jk_m(r_{l}^{(1)}\!-\!r_{l})},\cdots\!, e^{-jk_m(r_{l}^{(N)}\!-\!r_{l})}]^\mathrm{T}\!.
\end{align}
In particular, for the reflected path, the distance is calculated by $r_{l}={r}_{{l},f_1}+{r}_{{l},f_2}$, whereas for scattering paths, it is determined by $r_{l}={r}_{{l},s_2}$. By applying the second-order Taylor series expansion, the distance to the $n$-th antenna, ${r_{l}^{(n)}}$, is approximated by
$r_{l}^{(n)}\!\approx\!r_{l}\!-\!\delta_nd\sin\vartheta_{l}+\frac{\delta_n^2d^2(1-\sin^2\vartheta_{l})}{2r_{l}}$, ${\delta_{n}\!\!=\!\!\frac{2n-N-1}{2}}$, ${n\in\{1,\ldots, N\}}$. Here, ${\vartheta_{l}}$ represents the physical angle of the user's image relative to the reference antenna in the reflected path, and the physical angle of the last scatterer relative to the reference antenna in the scattering path, as illustrated in Fig.~\ref{Figure_2}.
\color{black}
\par Owing to the limited transmission paths in mmWave, the inherent sparsity in the polar-domain can be potentially exploited to achieve near-field channel extrapolation with a reduced number of measurements \cite{cema_zhang}, \cite{outlook_lei}. In near-field scenarios, the channel can be transformed into the polar-domain through a polar-domain dictionary ${\mathbf{W}}$. As discussed in \cite{wideband_dai}, ${\mathbf{W}\in\mathbb{C}^{N\times N_{a}N_{d}}}$ is composed of $N_{d}$ submatrices, where ${\left.\mathbf{W}=\left[\begin{matrix}{\mathbf{W}_{1},\mathbf{W}_{2},\cdots,\mathbf{W}_{N_{d}}}\\\end{matrix}\right.\right]}$. Each submatrix $\mathbf{W}_{n_d}\in\mathbb{C}^{N\times N_{a}}$, $n_d\in\{1,2,\cdots,N_d\}$, contains $N_a$ array steering vectors at the carrier frequency $f_c$, and can be represented as $\mathbf{W}_{n_d}=[\boldsymbol{b}(\overline{\vartheta}_1,\overline{r}_{1,n_d},f_c),\cdots,$ $\boldsymbol{b}(\overline{\vartheta}_{N_a},\overline{r}_{N_a,n_d},f_c)]$. Among them, the on-grid angles are uniformly sampled such that $\sin\overline{\vartheta}_{n_a}=\frac{2(n_a-1)-N_a}{N_a}$, $n_a\in\{1,2,\cdots,N_a\}$, while on-grid distances are non-uniformly sampled as $\overline{r}_{n_a,n_d}=\frac{D^2\cos^2\overline{\vartheta}_{n_a}}{2\beta^2\lambda_cn_d}$, where $\beta$ is a predefined parameter\footnote{The predefined parameter governs the correlation between near-field steering vectors, which determines the reconstruction accuracy.}. Then, (1) can be equivalently restated as
\vspace{-0.5em}\begin{align}
\mathbf{h}_m={\mathbf{W}}\mathbf{h}_p(m),
\end{align}
where ${\mathbf{h}_p(m)\in\mathbb{C}^{N_aN_d\times1}}$ represents the polar-domain channel at the $m$-th subcarrier. 
\par Considering the uplink transmission with a single pilot ${x}_{m}\in\mathbb{C}$, we denote the received signal at the $m$-th subcarrier as ${\mathbf{r}_{m}\in\mathbb{C}^{N\times 1}}$ at the BS, which can be denoted by
\vspace{-0.5em}\begin{align}
\mathbf{r}_{m}=\mathbf{h}_m{x}_{m}+\tilde{\mathbf{n}}_{m}.
\end{align}
In (4), the complex Gaussian noise ${\tilde{\mathbf{n}}_{m}\in\mathbb{C}^{N\times1}}$ follows the distribution ${\mathcal{CN}(\mathbf{0},\sigma^{2}\mathbf{I}_{N})}$, where $\sigma^{2}$ denotes the noise power at each received antenna. To further introduce the near-field spatial-domain channel extrapolation\footnote{To focus on the spatial-domain channel extrapolation task, the effect of phase shifters is deliberately neglected during the CSI acquisition phase.}, we define the partially received signal ${\mathbf{y}_{m}\in\mathbb{C}^{K\times1}}$ on the $m$-th subcarrier as
\vspace{-0.5em}\begin{align}
\begin{aligned}\mathbf{y}_{m}&=\mathbf{A}\mathbf{r}_{m}=\mathbf{A}(\mathbf{h}_m{x}_{m}+\tilde{\mathbf{n}}_{m}).\end{aligned}
\end{align}
\color{black}
The binary measurement matrix ${\mathbf{A}\in\mathbb{R}^{K\times N}}$ is determined by the pattern of antenna selection $\mathbf{\Xi} \in\mathbb{R}^{K\times1}$, which can be formulated as
\vspace{-0.5em}\begin{align}
  \left.\left[\mathbf{A}\right]_{k,n}=\left\{
\begin{array}
{cc}1, & n\in\mathbf{\Xi}\left(k\right); \\
0, & \mathrm{else}.
\end{array}\right.\right.
\end{align}\vspace{-0.5em}
\par For simplicity and without loss of generality, we assume that the pilot symbol ${x_{m}\!=\!1}$ and then the post-processed baseband signal can be rewritten as
\vspace{-0.5em}\begin{align}
\mathbf{y}_{m}=\mathbf{A}\mathbf{h}_m+\mathbf{n}_{m},
\end{align}
where ${\mathbf{n}_{m}=\mathbf{A}\tilde{\mathbf{n}}_{m}\in\mathbb{C}^{K\times1}}$. Based on (3), we can re-express (7) as
\vspace{-0.5em}\begin{align}
\begin{aligned}\mathbf{y}_{m} &=\mathbf{A}\mathbf{W}\mathbf{h}_{p}(m)+\mathbf{n}_{m}\\
&=\mathbf{\Psi}\mathbf{h}_{p}(m)+\mathbf{n}_{m},\end{aligned}
\end{align}
where the sensing matrix $\mathbf{\Psi}\in\mathbb{C}^{K\times{N_aN_d}}$ is defined by $\mathbf{\Psi}=\mathbf{A}\mathbf{W}$. Moreover, after transmitting over all $M$ subcarriers, the received signal can be expressed as
\vspace{-0.5em}\begin{align}
\mathbf{Y}=\mathbf{A}\mathbf{H}+\mathbf{N}=\mathbf{\Psi}\mathbf{H}_p+\mathbf{N},
\end{align}
where $\mathbf{Y}=[\mathbf{y}_1,\mathbf{y}_2,\dots,\mathbf{y}_M]\in\mathbb{C}^{K\times{M}}$, $\mathbf{H}=[\mathbf{h}_1$, $\mathbf{h}_2, \dots, \mathbf{h}_M]\in\mathbb{C}^{{N}\times{M}}$, $\mathbf{H}_p=[\mathbf{h}_p(1),\mathbf{h}_p(2),\dots,$ $\mathbf{h}_p(M)]\in\mathbb{C}^{N_aN_d\times{M}}$, and $\mathbf{N}=[\mathbf{n}_1,\mathbf{n}_2, \dots, \mathbf{n}_M]\in\mathbb{C}^{K\times{M}}$.

\vspace{-1em}
\section{Proposed Framework for near-field spatial-domain Channel Extrapolation}\label{se:model}
\par In Section III-A, we formulate an optimization problem employing a multi-subcarrier joint scheme. Then, in Section III-B, we propose an on-grid adaptive spatial-domain channel extrapolation algorithm, termed P-ASOMP, within the MMV framework. To further enhance accuracy, we introduce an off-grid adaptive channel extrapolation algorithm, named P-ASIGW. Additionally, to mitigate the high computational complexity, we propose a CV-based scheme within the MMV framework in Section III-C, enabling low-complexity near-field channel extrapolation.
\vspace{-1.5em}
\subsection{Problem Formulation}
 Let $\mathcal{A}\in\mathbb{R}^{N\times1}$ denote the complete set of all antennas and $\mathcal{B}\in\mathbb{R}^{K\times1}$ represent a subset of $\mathcal{A}$. Accordingly, we define $\mathbf{H}^\mathcal{A}\in\mathbb{C}^{N\times{M}}$ (equivalent to $\mathbf{H}\in\mathbb{C}^{N\times{M}}$ in (9)) as the comprehensive channel matrix encompassing all antennas, and $\mathbf{H}^\mathcal{B}\in\mathbb{C}^{K\times{M}}$ (equivalent to $\mathbf{Y}\in\mathbb{C}^{K\times{M}}$ in (9)) as the partially received signal from the channel corresponding to the subset $\mathcal{B}$. For a given static communication environment, the bijectivity of the position-to-channel mapping function has been demonstrated in \cite{mapping_alra}. Consequently, a channel-to-channel mapping is established \cite{8e_gao}. This correspondence makes it feasible to extrapolate the full CSI from the partially received signal $\mathbf{H}^\mathcal{B}\in\mathbb{C}^{K\times{M}}$. As illustrated in Fig.~\ref{Figure_1}, the extrapolated channel is defined by $\hat{\mathbf{H}}^\mathcal{A}\in\mathbb{C}^{N\times{M}}$. Given that the extrapolation problem inherently involves reconstructing a high-dimensional signal from limited observations, we formulate it within the CS framework. By leveraging CS theory, the high-dimensional matrix recovery problem is formulated as an $l_0$-norm minimization to identify the sparsest solution. However, due to its nondeterministic polynomial-time (NP)-hard nature \cite{mathematical}, a common alternative is to employ $l_1$-norm minimization, as detailed in \cite{l1_candes}. Under the MMV framework, we further reformulate the sparse signal recovery problem with a row-wise $l_{2,1}$ norm minimization as follows \cite{adaptive_zhang}:
\vspace{-0.5em}\begin{align}
\hat{\mathbf{H}}_p\!=\!\arg\min\!\left\|\mathbf{H}_p\right\|_{2,1},\mathrm{s.t.}\left\|\mathbf{H}^\mathcal{B}\!-\!\mathbf{\Psi H}_p\right\|_\mathrm{F}\!\leq\!\epsilon,
\end{align}
where $\epsilon$ is explicitly determined by the noise level\footnote{To ensure accurate recovery, an upper bound on the residual power has been verified in \cite{hierarchical_cai}, \cite{CS_bound}. In this paper, we continue to adopt this limit as the halting condition.} and $\hat{\mathbf{H}}_p\in\mathbb{C}^{N_aN_d\times{M}}$ is the estimate of $\mathbf{H}_p\in\mathbb{C}^{N_aN_d\times{M}}$. Importantly, unlike conventional hybrid architectures, the sensing matrix $\mathbf{\Psi}$ in the spatial-domain channel extrapolation problem is determined by the antenna selection pattern rather than the analog combining network.
\color{black}
Then, the extrapolated near-field channel can be denoted as 
\vspace{-0.5em}\begin{align}
  \hat{\mathbf{H}}^\mathcal{A}=\mathbf{W}\hat{\mathbf{H}}_p.
\end{align}
As shown in (11), the complete near-field channel is reconstructed exploiting the polar-domain dictionary $\mathbf{W}$.
\vspace{-1.5em}
\subsection{Proposed Polar-domain Adaptive Channel Extrapolation Algorithms}
\vspace{-0.5em}
\par Accurately determining the exact number of multipath components in practical scenarios is inherently challenging and often incurs additional overhead. This uncertainty directly affects the appropriate choice of iteration count in CS-based algorithms. Specifically, adopting insufficient iterations may degrade reconstruction accuracy, while excessive iterations can lead to noise-fitting and increase computational complexity.
\par To address this issue, we investigate the coherence properties of spatial-domain channels by leveraging a polar-domain dictionary, which effectively captures the inherent angular- and distance-dependent structures of near-field propagation. Motivated by this insight, we propose an on-grid P-ASOMP algorithm with an adaptive halting condition to improve the extrapolation efficiency and robustness, as detailed in \textbf{Algorithm 1}.
\color{black}
\begin{algorithm}
\caption{Proposed P-ASOMP Algorithm}\label{alg:cap}
\begin{algorithmic}[1]
\Require Partially received signal ${\mathbf{H}}^\mathcal{B}
$, sensing matrix ${\mathbf{\Psi}}$, the threshold $\epsilon$.
\Ensure The complete channel ${\hat{\mathbf{H}}^\mathcal{A}}$, number of detected paths $\hat{L}$.
\State Initialize: ${\mathbf{R}={\mathbf{H}}^\mathcal{B}}$, ${\mathbf{\Gamma}=\{\phi\}}$, $l=0$.
\State \textbf{repeat}
\State $l=l+1$;
\State Correlation: ${\mathbf{E}={\mathbf{\Psi}}^\mathrm{H}\mathbf{R}=\bigl[\mathbf{e}_{1},\mathbf{e}_{2},\cdots,\mathbf{e}_{M}\bigr].}$
\State Detect new support: 
${p^{*}=\arg\max_{p}\sum_{m=1}^{M}\|\mathbf{e}_{m}\|^{2}}$.
\State Update support set: ${\mathbf{\Gamma}_l=\mathbf{\Gamma}_{l-1}\cup{p^{*}}}$.
\State Orthogonal projection: ${\hat{\mathbf{H}}_p={\mathbf{\Psi}}^{\dagger}(:,\mathbf{\Gamma}_l){\mathbf{H}}^\mathcal{B}}$.
\State Update residual matrix: ${\mathbf{R}_l={\mathbf{H}}^\mathcal{B}-{\mathbf{\Psi}}(:,\mathbf{\Gamma}_l)\hat{\mathbf{H}}_p}$.

\State \textbf{until} $\|\mathbf{R}_l\|_\mathrm{F} \leq \epsilon$
\State $\hat{\mathbf{H}}^\mathcal{A}=\mathbf{W}_{:,\mathbf{\Gamma}_l}\hat{\mathbf{H}}_{p}, \hat{L}=l.$
\end{algorithmic}
\end{algorithm}
\vspace{-1.5em}
\par We first initialize the residual matrix, the support set, and the number of paths in step 1. Subsequently, the adaptive near-field channel extrapolation process is carried out in steps 2-9. The halting condition is defined by the upper bound on the residual power, given as $\epsilon=\sigma\sqrt{KM}$. Then, in step 4, the projection coefficients of each column of the sensing matrix onto the residual matrix are computed\footnote{Since the center frequency $f_c$ substantially exceeds the system bandwidth, all subcarrier frequencies $f_m$ are approximated as equal to the center frequency, i.e., $f_m \approx f_c$. Consequently, during the detection process, it is assumed that all subcarriers share an identical support set. However, as the bandwidth increases, this approximation becomes invalid, and the beam-split effect becomes pronounced \cite{wideband_dai}. To address this issue, a frequency-dependent polar-domain dictionary can be designed to mitigate the resulting mismatch, as demonstrated in \cite{adaptive_zhang}.}, effectively capturing the spatial correlation through the polar-domain dictionary designed at the center frequency $f_c$ under limited measurements. In step 5, the physical location of the $l$-th component is determined by the index ${p^{*}=\arg\max_{p}\sum_{m=1}^{M}\|\mathbf{e}_{m}\|^{2}}$. We further add the index $p^{*}$ to the support set $\mathbf{\Gamma}$. Following this, the polar-domain channel is estimated using the least squares (LS)\footnote{In order to ensure that the residuals are orthogonal to the chosen atomic space, LS is widely adopted in existing work \cite{hybrid_zhang}, \cite{polar_dai}.} algorithm in step 7. Subsequently, in step 8, the residual matrix is updated by removing the estimated path components. Finally, the complete near-field channel is reconstructed leveraging the polar-domain dictionary in step 10.
\par Since the proposed P-ASOMP algorithm operates on a sampled polar-domain (i.e. on-grid angles and distances), its channel extrapolation accuracy is limited. To improve accuracy and better approximate the actual continuous distribution of angles and distances (i.e. off-grid angles and distances), we refine these parameters based on the outcome of the P-ASOMP algorithm. Inspired by the P-SIGW algorithm proposed in \cite{polar_dai}, we further introduce an off-grid P-ASIGW algorithm, which leverages the number of detected paths $\hat{L}$ from the P-ASOMP algorithm, as detailed in \textbf{Algorithm 2}.
\begin{algorithm}
\caption{Proposed P-ASIGW Algorithm}\label{alg:cap}
\begin{algorithmic}[1]
\Require Partially received signal ${\mathbf{H}}^\mathcal{B}
$, measurement matrix $\mathbf{A}$, the minimum distance $\rho_\mathrm{min}$, and number of iterations $N_\mathrm{iter}$.
\Ensure The complete channel ${\hat{\mathbf{H}}^\mathcal{A}}$.
\State Obtain the initial estimates of the angles $(\hat{\boldsymbol{\theta}}^{0})^\mathrm{T}=[\hat{\theta}_{1}^{0}, \hat{\theta}_{2}^{0}, \dots, \hat{\theta}_{\hat{L}}^{0}]$, the distances $(\hat{\mathbf{r}}^{0})^\mathrm{T}=[\hat{r}_{1}^{0}, \hat{r}_{2}^{0}, \dots, \hat{r}_{\hat{L}}^{0}]$, and the number of detected paths $\hat{L}$ by \textbf{Algorithm 1}.
\For {$n\in\{1, 2, \dots, N_\mathrm{iter}\}$}
\State Choose the strong Wolfe backtracking line search step length $l_1$.
\State Update the angles by (15).
\State Choose the strong Wolfe backtracking line search step length $l_2$.
\State Update the distances by (16).
\State Update the path gains by (13).
\EndFor
\State $\hat{\mathbf{H}}^\mathcal{A}=\left[ \mathbf{b} \left( \hat{\theta}_1^n, \hat{r}_1^n \right), \mathbf{b} \left( \hat{\theta}_2^n, \hat{r}_2^n \right), \cdots, \mathbf{b} \left( \hat{\theta}_{\hat{L}}^n, \hat{r}_{\hat{L}}^n \right) \right] \hat{\mathbf{G}}^n
.$
\end{algorithmic}
\end{algorithm}
\par The proposed P-ASIGW algorithm introduces significant innovations in both the initialization stage and refinement stage compared to the P-SIGW algorithm. During the initialization stage, we determine the initial value of the angles $(\hat{\boldsymbol{\theta}}^{0})^\mathrm{T}=[\hat{\theta}_{1}^{0}, \hat{\theta}_{2}^{0}, \dots, \hat{\theta}_{\hat{L}}^{0}]\in\mathbb{C}^{1\times{\hat{L}}}$, the distances $(\hat{\mathbf{r}}^{0})^\mathrm{T}=[\hat{r}_{1}^{0}, \hat{r}_{2}^{0}, \dots, \hat{r}_{\hat{L}}^{0}]\in\mathbb{C}^{1\times{\hat{L}}}$, and the number of detected paths $\hat{L}$ by executing \textbf{Algorithm 1}. Owing to the adaptive halting condition, the reliance on the actual number of path components is eliminated, which enhances usability and avoids excessive iterations. To maximize the likelihood $(\hat{\boldsymbol{\theta}},\hat{\mathbf{r}},\hat{\mathbf{G}})=\arg\max_{\boldsymbol{\theta},\mathbf{r},\mathbf{G}}p\left(\mathbf{H}^{\mathcal{B}}\mid\boldsymbol{\theta},\mathbf{r},\mathbf{G}\right)$, we alternatively optimize the angles $\hat{\boldsymbol{\theta}}^\mathrm{T}\in\mathbb{C}^{1\times{\hat{L}}}$, distances $\hat{\mathbf{r}}^\mathrm{T}\in\mathbb{C}^{1\times{\hat{L}}}$, and path gains $\hat{\mathbf{G}}\in\mathbb{C}^{N_aN_d\times{M}}$, which is formulated as 
\vspace{-0.5em}\begin{align}
\min_{\hat{\mathbf{G}},\hat{\boldsymbol{\theta}},\hat{\mathbf{r}}}\|{\mathbf{H}}^\mathcal{B}-\tilde{{\boldsymbol{\Psi}}}(\hat{\boldsymbol{\theta}},\hat{\mathbf{r}})\hat{\mathbf{G}}\|_{F}^{2}.
\end{align}
Since the optimization problem in (12) is non-convex \cite{polar_dai}, we utilize the alternating minimization method to address it. Then, the optimal solution for $\hat{\mathbf{G}}^\mathrm{opt}\in\mathbb{C}^{N_aN_d\times{M}}$ can be given by
\vspace{-0.5em}\begin{align}
\hat{\mathbf{G}}^\mathrm{opt}=\tilde{{\boldsymbol{\Psi}}}^\dagger(\hat{\boldsymbol{\theta}},\hat{\mathbf{r}}){\mathbf{H}}^\mathcal{B}.
\end{align}
Substitute (13) into (12), the maximum-likelihood problem is reformulated as
\vspace{-0.5em}\begin{align}
\begin{aligned}
&\min_{\hat{\boldsymbol{\theta}},\hat{\mathbf{r}}} \|{\mathbf{H}}^\mathcal{B}-\tilde{{\boldsymbol{\Psi}}}(\hat{\boldsymbol{\theta}},\hat{\mathbf{r}})\tilde{{\boldsymbol{\Psi}}}^\dagger(\hat{\boldsymbol{\theta}},\hat{\mathbf{r}}){\mathbf{H}}^\mathcal{B}\|_F^2 \\
&\iff \min_{\hat{\boldsymbol{\theta}},\hat{\mathbf{r}}} \mathrm{Tr}\left\{{{\mathbf{H}}^\mathcal{B}}^\mathrm{H}\left(\mathbf{I}-\mathbf{P}(\hat{\boldsymbol{\theta}},\hat{\mathbf{r}})\right)^\mathrm{H}\left(\mathbf{I}-\mathbf{P}(\hat{\boldsymbol{\theta}},\hat{\mathbf{r}})\right){\mathbf{H}}^\mathcal{B}\right\}\\
&\overset{(a)}{\iff}\min_{\hat{\boldsymbol{\theta}},\hat{\mathbf{r}}}\mathcal{L}(\hat{\boldsymbol{\theta}},\hat{\mathbf{r}})=-\mathrm{Tr}\left\{{{\mathbf{H}}^\mathcal{B}}^\mathrm{H}\mathbf{P}(\hat{\boldsymbol{\theta}},\hat{\mathbf{r}}){\mathbf{H}}^\mathcal{B}\right\},
\end{aligned}
\end{align}
where $\mathbf{P}(\hat{\boldsymbol{\theta}},\hat{\mathbf{r}})=\tilde{{\boldsymbol{\Psi}}}(\hat{\boldsymbol{\theta}},\hat{\mathbf{r}})\tilde{{\boldsymbol{\Psi}}}^\dagger(\hat{\boldsymbol{\theta}},\hat{\mathbf{r}})\in\mathbb{C}^{K\times{K}}$ and (a) is satisfied by $\mathbf{P}^\mathrm{H}(\hat{\boldsymbol{\theta}},\hat{\mathbf{r}})\mathbf{P}(\hat{\boldsymbol{\theta}},\hat{\mathbf{r}})=\mathbf{P}(\hat{\boldsymbol{\theta}},\hat{\mathbf{r}})$. Then, an iterative gradient descent method\footnote{The details of gradient $\nabla\mathcal{L}(\hat{\boldsymbol{\theta}},\hat{\mathbf{r}})$ can be found in Appendix B of \cite{polar_dai}.} is utilized to optimize $\hat{\boldsymbol{\theta}}$ and $\hat{\mathbf{r}}$ with respect to the new objective function $\mathcal{L}(\hat{\boldsymbol{\theta}},\hat{\mathbf{r}})$. In the $n$-th iteration of the alternating optimization process, the angles are updated as
\vspace{-0.5em}\begin{align}
\hat{\boldsymbol{\theta}}^n=\hat{\boldsymbol{\theta}}^{n-1}-l_1\nabla_{\hat{\boldsymbol{\theta}}}\mathcal{L}(\hat{\boldsymbol{\theta}},\hat{\mathbf{r}}^{n-1})|_{\hat{\boldsymbol{\theta}}=\hat{\boldsymbol{\theta}}^{n-1}},
\end{align}
where $l_1$ denotes the step length for the angle search. Since the distance-sampling method in the polar-domain dictionary is non-uniform for $r$ but uniform for $\frac{1}{r}$, the distances are updated with respect to $\frac{1}{\hat{\mathbf{r}}}=[\frac{1}{\hat{r}_1},\frac{1}{\hat{r}_2},\dots,\frac{1}{\hat{r}_{\hat{L}}}]$ to indirectly update $\hat{\mathbf{r}}$, ensuring stable performance. In the $n$-th iteration, the distances are updated as
\vspace{-0.5em}\begin{align}
\frac{1}{\hat{\mathbf{r}}^{n}}=\frac{1}{\hat{\mathbf{r}}^{n-1}}-l_{2}\nabla_{\frac{1}{\hat{\mathbf{r}}}}\mathcal{L}(\hat{\boldsymbol{\theta}}^{n},\hat{\mathbf{r}})|_{\hat{\mathbf{r}}=\hat{\mathbf{r}}^{n-1}},
\end{align}
where $l_2$ denotes the step length for the distance search. To ensure the non-increasing property, the P-SIGW algorithm updates the step lengths $l_1$ and $l_2$ leveraging Armijo backtracking line search \cite{polar_dai}. Furthermore, to accelerate convergence, strong Wolfe conditions\footnote{When the array size is not extremely large, the strong Wolfe condition leads to faster convergence. However, as the array size increases substantially, its stricter curvature requirement may compromise stability; thus, adopting a relaxed Armijo rule is more effective.} are applied in this paper. Taking the angle search as an example, these conditions include two main components: the sufficient decrease condition, which aligns with the Armijo condition in (17), and the curvature condition, as outlined in (18):
\vspace{-0.5em}\begin{align}
\mathcal{L}(\hat{\boldsymbol{\theta}}^n)\leq\mathcal{L}(\hat{\boldsymbol{\theta}}^{n-1})+c_1l_1\nabla_{\hat{\boldsymbol{\theta}}}\mathcal{L}(\hat{\boldsymbol{\theta}}^{n-1})^\mathrm{T}d^n,
\end{align}\vspace{-0.5em}
\vspace{-0.5em}\begin{align}
|\nabla_{\hat{\boldsymbol{\theta}}}\mathcal{L}(\hat{\boldsymbol{\theta}}^n)^\mathrm{T}d^n|\leq c_2|\nabla_{\hat{\boldsymbol{\theta}}}\mathcal{L}(\hat{\boldsymbol{\theta}}^{n-1})^\mathrm{T}d^n|,
\end{align}
where $d^n=-\nabla_{\hat{\boldsymbol{\theta}}}\mathcal{L}(\hat{\boldsymbol{\theta}}^{n-1})$, $c_{1}\in(0,\frac{1}{2})$, and $c_{2}\in(c_{1},1)$. 
Based on (17) and (18), the step lengths for the angle search are determined in step 3. As in step 4, the angles are updated adopting (15). Similarly, the step lengths for the distance search are selected in step 5, and the distances are updated using (16) in step 6. Then, the path gains are updated according to (13).
\par Finally, after $n$ iterations, the updated near-field path components are exploited to construct the matrix $\tilde{\mathbf{W}}(\hat{\boldsymbol{\theta}},\hat{\mathbf{r}})=\left[ \mathbf{b} \left( \hat{\theta}_1^n, \hat{r}_1^n \right), \mathbf{b} \left( \hat{\theta}_2^n, \hat{r}_2^n \right), \cdots, \mathbf{b} \left( \hat{\theta}_{\hat{L}}^n, \hat{r}_{\hat{L}}^n \right) \right]$, and then the complete near-field channel can be reconstructed as $\hat{\mathbf{H}}^\mathcal{A}=\tilde{\mathbf{W}}(\hat{\boldsymbol{\theta}},\hat{\mathbf{r}})\hat{\mathbf{G}}^n$ in step 9.
\subsection{Proposed Cross-validation based Polar-domain Adaptive Channel Extrapolation Algorithms}
\par In XL-MIMO systems, the large number of antennas and the near-field effect jointly impose high computational complexity. Although parameter decoupling has been proven effective for complexity reduction in related work, e.g., \cite{2D_chau}, its application to channel extrapolation remains constricted due to the challenge of estimating the array covariance matrix from limited observations, thereby necessitating novel algorithmic solutions.
\par Notably, the correlation step in the OMP framework represents a major source of computational complexity, restricting the practicality of adaptive extrapolation algorithms such as those introduced in Section III-B. To mitigate this limitation, we propose a novel CV-based scheme, inspired by the common support set assumption in the MMV framework. Specifically, the goal of this scheme is to accurately identify the common support set by adopting only a minimal subset of training data. The detailed procedure of the resulting on-grid algorithm, termed CV-P-ASOMP, is presented in \textbf{Algorithm 3}.
\color{black}
\par Exploiting the common support set assumption across subcarriers in the MMV framework, our method strategically partitions them into two distinct groups: (1) a training set for support set identification; (2) a validation set for evaluating the termination criterion. In step 1, the CV parameters are determined by the CV ratio $\alpha$, yielding $T\!=\!\lceil \alpha M \rceil$ training subcarriers and $V\!=\!M\!-\!T$ validation subcarriers. To ensure coverage across the entire bandwidth, the training set is selected via uniform sampling, i.e., $\mathcal{I}_{\text{train}}\!=\![1,1\!+\!\Delta,\dots,1\!+\!(T\!-\!1)\Delta]$, where the sampling interval is set as $\Delta\!=\!\left\lfloor \frac{M}{T} \right\rfloor$. The validation set is then defined as $\mathcal{I}_{\text{val}}\!=\!\mathcal{I}_{\text{all}}\setminus\mathcal{I}_{\text{train}}$, with $\mathcal{I}_{\text{all}}=[1,\dots,M]$. During initialization, we divide the residual matrix into two parts: the training residual matrix ${\mathbf{R}_{\mathrm{T}}\!=\!{\mathbf{H}}^\mathcal{B}_{:,\mathcal{I}_{\text{train}}}}\in\mathbb{C}^{K\times{T}}$ and the validating residual matrix  ${\mathbf{R}_{\mathrm{V}}\!=\!{\mathbf{H}}^\mathcal{B}_{:,\mathcal{I}_{\text{val}}}}\in\mathbb{C}^{K\times{V}}$. Next, the adaptive simultaneous orthogonal matching pursuit is executed in steps 2-10, with the same stopping criterion as in \textbf{Algorithm 1}. Notably, the correlation operation is computed only exploiting the training residual matrix ${\mathbf{R}_{\mathrm{T}}\!=\!{\mathbf{H}}^\mathcal{B}_{:,\mathcal{I}_{\text{train}}}}\in\mathbb{C}^{K\times{T}}$, significantly reducing computational complexity. The support set for the $l$-th component is identified in step 6 and concatenated in step 7. Following the OMP framework, the polar-domain gains are calculated through the LS method \cite{polar_dai}. Subsequently, the residual matrices are updated by subtracting the estimated path components in step 9. Finally, in Step 11, the complete near-field channel is extrapolated by leveraging the polar-domain dictionary, which effectively captures the underlying spatial correlation structure. 
\vspace{-1em}
\begin{algorithm}
\caption{Proposed CV-P-ASOMP Algorithm}\label{alg:cap}
\begin{algorithmic}[1]
\Require Partially received pilot ${\mathbf{H}}^\mathcal{B}$, sensing matrix ${\mathbf{\Psi}}$, the threshold $\epsilon$, the ratio of cross-validation $\alpha$.
\Ensure The complete channel ${\hat{\mathbf{H}}^\mathcal{A}}$, number of detected paths $\hat{L}$.
\State Calculate the cross-validation parameters: $T\!=\!\lceil \alpha{M} \rceil$, $\Delta\!=\!\left\lfloor \frac{M}{T} \right\rfloor$, $\mathcal{I}_{\text{train}}\!=\![1,1\!+\!\Delta,\dots,1\!+\!(T\!-\!1)\Delta]$, $\mathcal{I}_{\text{val}}\!=\!\mathcal{I}_{\text{all}}\setminus\mathcal{I}_{\text{train}}$.
\State Initialize: ${\mathbf{R}_{\mathrm{T}}\!=\!{\mathbf{H}}^\mathcal{B}_{:,\mathcal{I}_{\text{train}}}}$, ${\mathbf{R}_{\mathrm{V}}\!=\!{\mathbf{H}}^\mathcal{B}_{:,\mathcal{I}_{\text{val}}}}$, ${\mathbf{\Gamma}=\{\phi\}}$, $l=0$.
\State \textbf{repeat}
\State $l=l+1$;
\State Correlation: ${\mathbf{E}={\mathbf{\Psi}}^\mathrm{H}\mathbf{R}_\mathrm{T}=\bigl[\mathbf{e}_{1},\mathbf{e}_{2},\cdots,\mathbf{e}_{T}\bigr].}$
\State Detect new support: 
${p^{*}=\arg\max_{p}\sum_{v=1}^{V}\|\mathbf{e}_{v}\|^{2}}$.
\State Update support set: ${\mathbf{\Gamma}_l=\mathbf{\Gamma}_{l-1}\cup{p^{*}}}$,
\State Orthogonal projection: ${\hat{\mathbf{H}}_l={\mathbf{\Psi}}^{\dagger}(:,\mathbf{\Gamma}_l){\mathbf{H}}^\mathcal{B}_{:,\mathcal{I}_{\text{train}}}}$, ${\Tilde{\mathbf{H}}_l={\mathbf{\Psi}}^{\dagger}(:,\mathbf{\Gamma}_l){\mathbf{H}}^\mathcal{B}_{:,\mathcal{I}_{\text{val}}}}$, ${\hat{\mathbf{H}}_p={\mathbf{\Psi}}^{\dagger}(:,\mathbf{\Gamma}_l){\mathbf{H}}^\mathcal{B}}$.
\State Update residual matrices: ${\mathbf{R}_\mathrm{T}={\mathbf{H}}^\mathcal{B}_{:,\mathcal{I}_{\text{train}}}-{\mathbf{\Psi}}(:,\mathbf{\Gamma}_l)\hat{\mathbf{H}}_l}$, ${\mathbf{R}_\mathrm{V}={\mathbf{H}}^\mathcal{B}_{:,\mathcal{I}_{\text{val}}}-{\mathbf{\Psi}}(:,\mathbf{\Gamma}_l)\Tilde{\mathbf{H}}_l}$.

\State \textbf{until} $\|\mathbf{R}_\mathrm{V}\|_\mathrm{F} \leq \epsilon$
\State $\hat{\mathbf{H}}^\mathcal{A}=\mathbf{W}_{:,\mathbf{\Gamma}_l}\hat{\mathbf{H}}_{p}, \hat{L}=l.$
\end{algorithmic}
\end{algorithm}
\vspace{-1em}
\par To address the error introduced by on-grid sample points, we propose the CV-P-ASIGW algorithm. Similar to \textbf{Algorithm 2}, the step lengths are chosen based on the strong Wolfe conditions; however, the key difference is that the initialization information derived from \textbf{Algorithm 3}. The off-grid CV-P-ASIGW algorithm is detailed in \textbf{Algorithm 4}.
\begin{algorithm}
\caption{Proposed CV-P-ASIGW Algorithm}\label{alg:cap}
\begin{algorithmic}[1]
\Require Partially received signal ${\mathbf{H}}^\mathcal{B}
$, measurement matrix $\mathbf{A}$, the minimum distance $\rho_\mathrm{min}$, and number of iterations $N_\mathrm{iter}$.
\Ensure The complete channel ${\hat{\mathbf{H}}^\mathcal{A}}$.
\State Obtain the initial estimates of the angles $(\hat{\boldsymbol{\theta}}^{0})^\mathrm{T}=[\hat{\theta}_{1}^{0}, \hat{\theta}_{2}^{0}, \dots, \hat{\theta}_{\hat{L}}^{0}]$, the distances $(\hat{\mathbf{r}}^{0})^\mathrm{T}=[\hat{r}_{1}^{0}, \hat{r}_{2}^{0}, \dots, \hat{r}_{\hat{L}}^{0}]$, and the number of detected paths $\hat{L}$ by \textbf{Algorithm 3}.
\For {$n\in\{1, 2, \dots, N_\mathrm{iter}\}$}
\State Choose the strong Wolfe backtracking line search step length $l_1$.
\State Update the angles by (15).
\State Choose the strong Wolfe backtracking line search step length $l_2$.
\State Update the distances by (16).
\State Update the path gains by (13).
\EndFor
\State $\hat{\mathbf{H}}^\mathcal{A}=\left[ \mathbf{b} \left( \hat{\theta}_1^n, \hat{r}_1^n \right), \mathbf{b} \left( \hat{\theta}_2^n, \hat{r}_2^n \right), \cdots, \mathbf{b} \left( \hat{\theta}_{\hat{L}}^n, \hat{r}_{\hat{L}}^n \right) \right] \hat{\mathbf{G}}^n
.$
\end{algorithmic}
\end{algorithm}
\par The main difference between the CV-P-ASIGW and P-ASIGW algorithms lies in the initialization strategy. Specifically, in the CV-P-ASIGW algorithm, the initial estimates of the angles $(\hat{\boldsymbol{\theta}}^{0} )^\mathrm{T}= [\hat{\theta}_{1}^{0}, \hat{\theta}_{2}^{0}, \dots, \hat{\theta}_{\hat{L}}^{0}] \in \mathbb{C}^{1 \times \hat{L}}$, the distances $(\hat{\mathbf{r}}^{0})^\mathrm{T} = [\hat{r}_{1}^{0}, \hat{r}_{2}^{0}, \dots, \hat{r}_{\hat{L}}^{0}] \in \mathbb{C}^{1 \times \hat{L}}$, and the estimated number of paths $\hat{L}$ are all obtained from \textbf{Algorithm 3} in Step~1. Benefiting from the tailored adaptive iteration and line search mechanisms, the proposed approach enables efficient reconstruction of the entire near-field channel with reduced computational complexity, while ensuring satisfactory performance.
\vspace{-1.5em}
\section{Design of Patterns for Antenna Selection}\label{se:model}
\par During the channel extrapolation phase, a subset of antennas is strategically selected to receive the signals. However, the primary challenge under this framework lies in determining the optimal antenna selection strategy from the entire array \cite{TCN_zhang}, \cite{GNN_zhang}. In this section, we address this problem and aim to design a better pattern for effective extrapolation.
\par For convenience, we define the compression rate as $\eta=\frac{N}{K}$, where $K$ is the number of selected antenna elements. The partially received signal $\mathbf{H}^\mathcal{B}\in\mathbb{C}^{K\times{M}}$ is constructed from the set of indices corresponding to the ULA elements covered by the pattern, as follows:
\vspace{-0.5em}\begin{align}
  \mathbf{\Xi}=\left\{\xi_0,\xi_1,\ldots,\xi_{\frac{N}{\eta}-1}\right\},
\end{align}\vspace{-0.5em}
where $0\leq\xi_{n}\leq N-1$ holds for $n\in\{0,1,\dots,\frac{N}{\eta}-1\}$. 
\color{black}
\vspace{-1em}
\subsection{Deterministic-Position Setups}
Deterministic-position setups for spatial-domain channel extrapolation can be interpreted as intuitive sampling strategies across the entire array. Specifically, we consider two types of patterns, namely, the dense uniform pattern and the sparse comb pattern.
\par{\it{1) Dense uniform pattern \cite{multi_han}:}} In this pattern setup, a contiguous block of $\frac{N}{\eta}$ adjacent elements in the ULA is selected. Without loss of generality, we assume that the starting index of the dense uniform pattern is $i$, where $i\in\{0,\dots,\eta-1\}$. Thus, the dense uniform pattern includes the following elements:
\vspace{-0.5em}\begin{align}
  \mathbf{\Xi}_{\mathrm{du},i}=\left\{\frac{Ni}{\eta},\frac{Ni}{\eta}+1,\ldots,\frac{N(i+1)}{\eta}-1\right\}.
\end{align}\vspace{-0.5em}
\color{black}
\par{\it{2) Sparse comb pattern \cite{multi_han}:}} This pattern follows a comb-like structure, in which one antenna element is uniformly selected from every $\eta$ elements. Let $i$ denote the starting index of the sparse comb pattern. Then, the selected antenna elements can be represented as
\vspace{-0.5em}\begin{align}
  \mathbf{\Xi}_{\mathrm{sc}}=\left\{i,i+\eta,\ldots,i+N-\eta\right\}.
\end{align}
\vspace{-4em}
\subsection{Random-Position Setups}
In addition to deterministic-position setups with regularly spaced measurement positions, random-position setups can also be employed for antenna selection, analogous to the concept of random matrices in compressed sensing theory \cite{cema_zhang}. 
\par Here, we consider two random antenna selection setups: (i) the sparse random pattern; (ii) the coherence-minimizing-based random pattern. The former is commonly adopted in existing works \cite{multi_han}, \cite{near-reconstruction_han}, while the latter is specifically introduced to minimize the mutual coherence, thereby improving extrapolation performance.
\par{\it{1) Sparse random pattern \cite{near-reconstruction_han}:}} This pattern randomly selects $\frac{N}{\eta}$ elements from the array. Due to its stochastic nature, this selection pattern cannot be represented explicitly in a fixed format. Let $\mathbf{\Xi}_{\mathrm{sr}}$ denote the sparse random pattern with a size of $K$, where the elements are selected from $\gamma=[1,\cdots,N]$.
\par{\it{2) Coherence-minimizing-based random pattern:}} For compressed sensing-based methods, the design of the sensing matrix is critical, as it must satisfy the restricted isometry property (RIP) \cite{simplerip}, which is a fundamental condition that underpins the success of the technique. Typically, the sensing matrix is designed based on the measurement matrix and the sparse basis matrix. However, in the spatial-domain channel extrapolation scheme considered in (10), since the measurement matrix is constructed exclusively from only a subset of the antenna positions, the DoFs in the measurement matrix design are limited \cite{cema_zhang}. Consequently, conventional measurement matrices cannot be directly adopted. Additionally, verifying the RIP is particularly challenging for two main reasons. First, the RIP requires a bounded condition number for all submatrices, where the total number of submatrices grows combinatorially as $\mathrm{C}_{N}^{K}$. Second, computing the spectral norm of these matrices is generally non-trivial \cite{simplerip}. As an alternative, mutual coherence serves as a simple yet effective measure to evaluate the suitability of the sensing matrix \cite{cema_zhang}, \cite{mathematical}. It has been demonstrated that lower coherence is associated with improved reconstruction performance \cite{mathematical}. Specifically, the mutual coherence of the sensing matrix is defined as
\begin{align*}
\begin{aligned}
\mu(\mathbf{\Psi})&\triangleq\max_{q<p}|\mathbf{\Psi}_{:,q}^\mathrm{H}\mathbf{\Psi}_{:,p}|= \max_{q<p} |(\mathbf{A}\mathbf{w}_q)^\mathrm{H}(\mathbf{A}\mathbf{w}_p)| \\
&= \max_{q<p} \left| \mathbf{w}_q^\mathrm{H} \left( \sum_{k=1}^{K} \mathbf{a}_k \mathbf{a}_k^\mathrm{H} \right) \mathbf{w}_p \right| \\
&= \max_{q<p} \left| \sum_{k=1}^{K} \langle\mathbf{w}_q, \mathbf{a}_k\rangle \langle\mathbf{a}_k, \mathbf{w}_p\rangle \right| \leq K \cdot C_{max}^2,
\end{aligned} \tag{22}
\end{align*}
where $p=N_{a}(n_{d}-1)+n_{a}$, $q=N_{a}(n_d^{\prime}-1)+n_a^{\prime}$, $\mathbf{w}_q=\mathbf{W}_{:,q}$, $\mathbf{a}_k=\mathbf{A}_{k,:}^\mathrm{H}$ and $C_{max}=\max_{k,j} |\langle\mathbf{a}_k, \mathbf{w}_j\rangle|$. In (22), the third equality is due to $\mathbf{A}^\mathrm{H}\mathbf{A}=\sum_{k=1}^{K} \mathbf{a}_k \mathbf{a}_k^\mathrm{H}$. Moreover, the result in (22) resembles the concept of incoherence \cite{incoherence}, which ensures sufficient diversity and randomness in the measurements and avoids concentrating the observations on partial bases. Specifically, incoherence requires that the rows of the measurement matrix should not be able to sparsely represent the columns of the sparse basis matrix, and vice versa \cite{incoherence}. It should be noted that the mutual coherence is jointly determined by the composition of the polar-domain dictionary and the selected antenna indices. The former is influenced by the parameter $\beta$, while the latter can be optimized through the antenna selection strategy. With $\beta$ fixed, the key challenge in designing $\mathbf{A}$ with minimal $\mu$ is to determine the optimal selection of $K\!=\!\frac{N}{\eta}$ elements for measurement. Accordingly, by taking the polar-domain dictionary as the sparse basis matrix, we further formulate the optimization problem as in (23). To address this problem, we propose a coherence-minimizing-based random pattern generation procedure, as detailed in \textbf{Algorithm 5}.
\vspace{-1.5em}
\begin{figure*}
\normalsize
\setcounter{equation}{22} 
\begin{equation}
\label{*}
\begin{aligned}
\min_{n(\mathbf{\Xi}_{\mathrm{cmr}})=K}\!\!\!\mu(\mathbf{\Psi})&=\min_{n(\mathbf{\Xi}_{\mathrm{cmr}})=K}\max_{q<p}|\mathbf{\Psi}_{:,q}^\mathrm{H}\mathbf{\Psi}_{:,p}|= \max_{q<p} |(\mathbf{A}\mathbf{w}_q)^\mathrm{H}(\mathbf{A}\mathbf{w}_p)| \\
&=\min_{n(\mathbf{\Xi}_{\mathrm{cmr}})=K}\max_{q<p}\left|\sum_{n\in\mathbf{\Xi}}e^{jk_c\left(-nd\sin\theta_{n_a^{\prime}}+n^2d^2\frac{\left(1-\sin^2\theta_{n_a^{\prime}}\right)}{2r_{n_d^{\prime}}}\right)}e^{-jk_c\left(-nd\sin\theta_{n_a}+n^2d^2\frac{\left(1-\sin^2\theta_{n_a}\right)}{2r_{n_d}}\right)}\right| \\
&=\min_{n(\mathbf{\Xi}_{\mathrm{cmr}})=K}\max_{q<p}\left|\sum_{n\in\mathbf{\Xi}_{\mathrm{cmr}}}e^{-j k_c \left( n d \left(\sin \theta_{n_a'} - \sin \theta_{n_a}\right) + n^2 d^2 \left( \frac{1 - \sin^2 \theta_{n_a'}}{2 r_{n_d'}} - \frac{1 - \sin^2 \theta_{n_a}}{2 r_{n_d}} \right) \right)}\right|.
\end{aligned}
\end{equation}
\end{figure*}
\begin{algorithm}
\caption{Coherence-minimizing-based Random Pattern Generation}\label{alg:cap}
\begin{algorithmic}[1]
\Require $\gamma = [1,\cdots,N]$, $\mathbf{W}$.
\Ensure ${\mathbf{\Xi}_\mathrm{cmr}}$.
\State Randomly generate $R$ patterns with length $K$ from $\gamma$.
\For{${r\in\{1,2,\cdots,R\}}$}
\For{$k= 1$ to $K$}
  \State Set $\mathbf{A}_r(k, {\mathbf{\Xi}_{r}(k)}) = 1$.\color{black}
\EndFor
\State Compute $\mathbf{\Psi}_r=\mathbf{A}_r\mathbf{W}$.
\State Calculate $\mu(\mathbf{\Psi}_r)$ according to (22).
\color{black}
\EndFor
\State Select ${\mathbf{\Xi}_\mathrm{cmr}}$ according to (23).
\end{algorithmic}
\end{algorithm}
\color{black}
\par Initially, the set of available indices is defined as $\gamma = [1,\cdots,N]$. The goal is to identify a sub-optimal random pattern of length $K=\frac{N}{\eta}$, selected from a total of $\mathrm{C}_{N}^{K}$ possible patterns. To ensure computational feasibility, we first generate $R$ random patterns, where $R$ is a positive integer satisfying $R \ll \mathrm{C}_{N}^{K}$. Subsequently, in steps 3-7, the coherence of the sensing matrix corresponding to each random pattern is computed. Finally, the random pattern ${\mathbf{\Xi}_\mathrm{cmr}}$ with the lowest coherence is selected from the $R$ candidate patterns. This approach effectively balances computational efficiency with the goal of minimizing coherence for improved extrapolation performance.
\vspace{-1.5em}
\subsection{Radiation Profile}
\begin{figure*}[t!]
\begin{minipage}[t]{0.6\linewidth}
\centering
\includegraphics[width=1.05\linewidth]{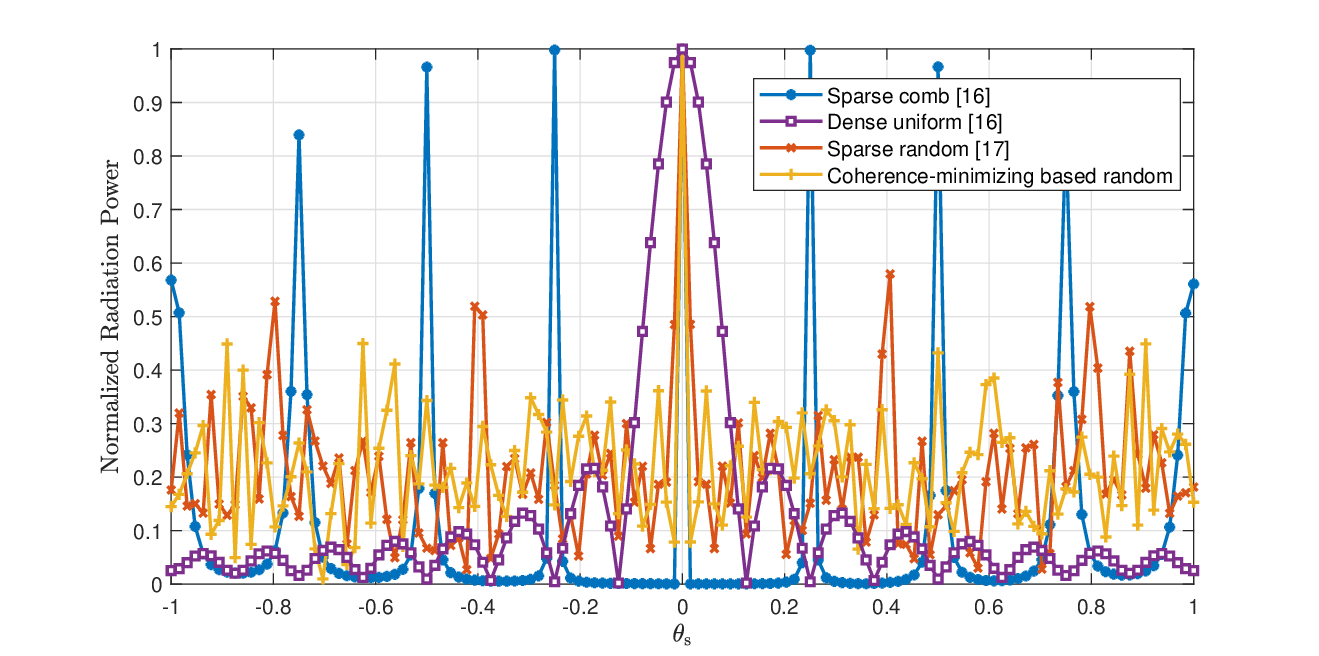}
\caption{\small{Radiation profiles of various pattern types with respect to $\theta_\mathrm{s}$.}\protect\label{Figure_3}}
\end{minipage}
\begin{minipage}[t]{0.4\linewidth}
\centering
\includegraphics[width=1.05\linewidth]{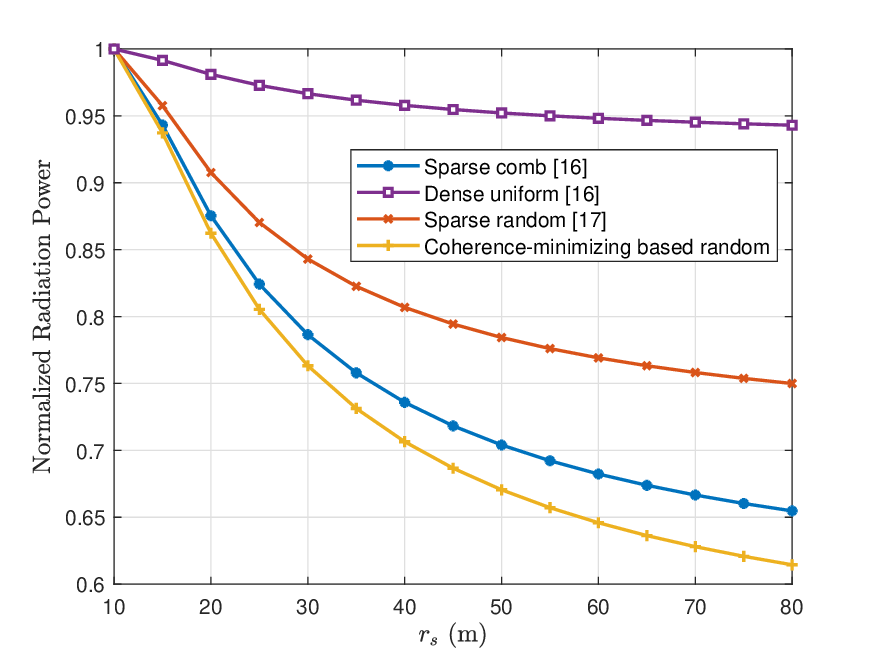}
\caption{\small{Radiation profiles of various pattern types with respect to $r_\mathrm{s}$.}\protect\label{Figure_4}}
\end{minipage}
\end{figure*}
To determine the most suitable pattern for channel extrapolation, we evaluate the performance of various pattern types by leveraging the radiation profile introduced in \cite{multi_han}. Specifically, the radiation power at any $\theta_{\mathrm{s}}$ and $r_{\mathrm{s}}$ is calculated as
\vspace{-0.5em}\begin{align}
P(\theta_{\mathrm{s}},r_{\mathrm{s}})\!=\!\frac{|\grave{\mathbf{a}}_c^\mathrm{H}(\theta_{\mathrm{s}},r_{\mathrm{s}}){\grave{\mathbf{a}}_c}(0,R_\mathrm{min})|}{\|{\grave{\mathbf{a}}_c}(\theta_{\mathrm{s}},r_{\mathrm{s}})\|},  \tag{24}
\end{align}
where we set $\theta=0$, $r=R_\mathrm{min}$ and design down-sampling steering vector ${\grave{\mathbf{a}}_c}({\theta}_{\mathrm{s}},r_{\mathrm{s}})$ based on frequency $f_c$. In Fig.~\ref{Figure_3} and Fig.~\ref{Figure_4}, we compare radiation profiles of four sampling patterns: the dense uniform pattern ${\mathbf{\Xi}_\mathrm{du}}$, sparse comb pattern ${\mathbf{\Xi}_\mathrm{sc}}$, sparse random pattern ${\mathbf{\Xi}_\mathrm{sr}}$, and coherence-minimizing-based random pattern ${\mathbf{\Xi}_\mathrm{cmr}}$, with a compression rate of $\eta=8$. We first examine the angular-domain radiation profile, assuming perfect knowledge of the distance. As shown in Fig.~\ref{Figure_3}, over the angular range ${\theta}_{\mathrm{s}} \in [-1, 1]$, a narrower main lobe centered at $\theta_{\mathrm{s}} = 0$ indicates superior angular resolution. Ideally, side lobes should be significantly lower than the main lobe to effectively mitigate false alarms in the presence of nearby paths \cite{multi_han}. Subsequently, Fig.~\ref{Figure_4} illustrates the distance-domain radiation profiles under the assumption that the angular direction of each sampling point is known a priori. Over the distance range $r_{\mathrm{s}} \in [10, 80]$, all patterns exhibit a distinct peak at the reference location with significantly reduced responses elsewhere. Importantly, a more rapid decay from the peak reflects enhanced distance resolution. This characteristic ensures that sampling points at other distances have minimal interference with the desired distance estimate.
\par{\it{1) Dense uniform pattern:}} It is evident that the dense uniform pattern exhibits a significantly wider main lobe compared to the other patterns, which is mainly determined by its aperture \cite{multi_han}. Moreover, this pattern achieves exceptionally low side lobe levels, rendering it highly effective at distinguishing the correct path in multipath environments. However, among the four patterns, it demonstrates the worst distance resolution. Nevertheless, as the number of active antennas increases, its distance resolution performance improves accordingly.
\par{\it{2) Sparse comb pattern:}} This pattern features a significantly narrower main lobe, precisely centered at $\theta_{\mathrm{s}}=0$. However, the compressed estimation range results in the emergence of $\eta$ side lobes \cite{multi_han}. While the majority of side lobes remain suppressed, the $\eta$ side lobes generate significant interference, complicating accurate identification of the desired propagation path. Moreover, unlike the far-field scenario, these side lobes exhibit varying levels of influence in the near-field, a difference attributed to the distinct modeling approaches for far-field and near-field channels \cite{polar_dai}. Additionally, the sparse comb pattern demonstrates better distance resolution performance compared to the dense uniform pattern. 
\par{\it{3) Sparse random pattern:}} We can clearly observe that this pattern exhibits a relatively narrow main lobe in the correct propagation direction, which is desirable for effective channel extrapolation. However, the maximum heights of its side lobes are higher than those of the dense uniform pattern. Additionally, due to its inherently random selection strategy, its distance resolution performance is unstable and generally worse than that of the sparse comb pattern.
\par{\it{4) Coherence-minimizing-based random pattern:}} By integrating the coherence-minimizing procedure in random pattern generation, this approach achieves a main lobe nearly as narrow as that of the sparse comb pattern, with its peak precisely at the target position $\theta_{\mathrm{s}}=0$. Furthermore, as illustrated in Fig.~\ref{Figure_4}, it exhibits the best distance resolution performance among the four considered patterns, where the normalized radiation power decays more rapidly than that of the other patterns and demonstrates more stable behavior compared to the sparse random pattern. From Fig.~\ref{Figure_3} and Fig.~\ref{Figure_4}, it is evident that the coherence-minimizing-based random pattern performs well in both angular and distance resolution, establishing it as a highly effective choice for channel extrapolation.
\color{black}
\vspace{-1em}
\section{Simulation Results}\label{se:model}
\par In this section, we evaluate the performance of our proposed near-field spatial-domain channel extrapolation algorithms through simulations under the CS framework. The simulation parameters are ${f_{c}=28 ~\mathrm{GHz}}$, ${B=28 ~\mathrm{MHz}}$, ${M=128}$, ${N=128}$, $U=3$. Therefore, the carrier wavelength is ${\lambda_{c}={\frac{c}{f_{c}}}=10.71 \mathrm{~mm}}$, the antenna spacing is ${d=\frac{\lambda_{c}}{2}=5.36 \mathrm{~mm}}$, and the BS array aperture is ${D=(N}$ ${-1)d\approx Nd=\frac{N\lambda_{c}}{2}=0.69 ~\mathrm{m}}$. Then, it can be calculated that the Rayleigh distance is ${\frac{2D^{2}}{\lambda_{c}}= 88.91\mathrm {~m}}$ \cite{frauhofer_KT}. 

The distances between the BS and near-field user or scatterers are randomly selected from $\mathcal{U}(10,80)$\footnote{In this paper, the Rayleigh distance is adopted as the boundary between the far-field and near-field regions. Accordingly, users located within the Rayleigh distance are treated as near-field users, and the study focuses on channel extrapolation by accounting for near-field characteristics. It is worth noting that other definitions for the near-field and far-field boundary also exist, as discussed in \cite{boundary_di}. Moreover, hybrid-field scenarios, as a more general and practical case, will be addressed in future work.}$\mathrm{~m}$.
\color{black}
In the design of the polar-domain dictionary, a parameter $\beta=1.5$ is employed to ensure adequate incoherence among near-field detection vectors. Throughout all simulation scenarios, the number of candidate patterns is set to $R=10$.
\color{black}
In this paper, the signal-to-noise ratio (SNR) is given by $\mathrm{SNR}=\frac{p_t}{\sigma^2}=\frac{1}{\sigma^2}$. 
\vspace{-1em}
\subsection{Pattern Designs Comparison}
\par In Section IV-C, we evaluate the performance of different pattern types adopting radiation profiles. 
Next, we evaluate four antenna selection patterns, dense uniform, sparse comb, sparse random, and coherence-minimizing-based random patterns, in the channel extrapolation problem to assess their effectiveness. For a fair comparison, the channel extrapolation corresponding to all patterns is implemented using \textbf{Algorithm 1}.
\begin{figure} [t!]
	\centering
	\subfloat[\label{1a}]{
		\includegraphics[scale=0.5]{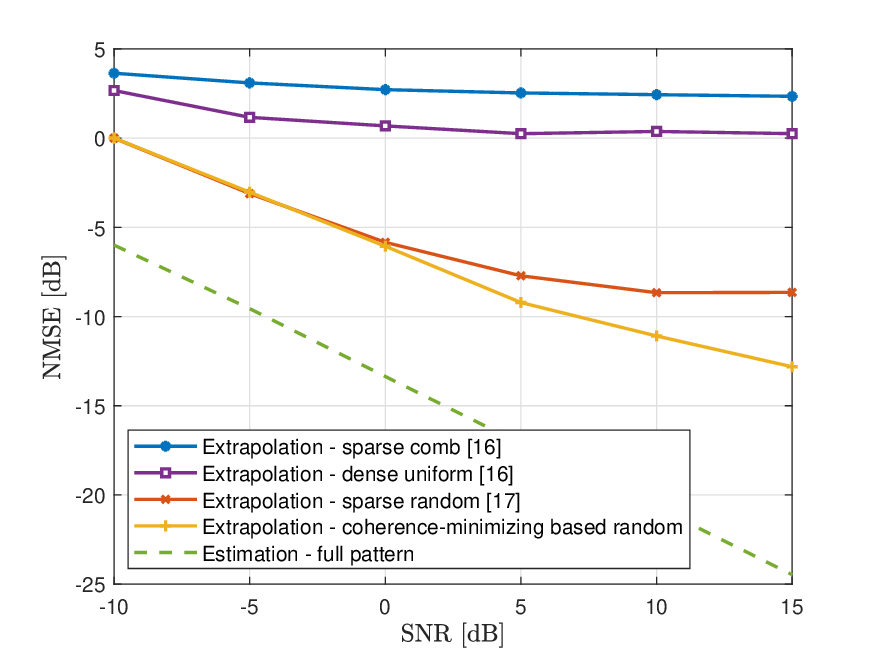}} \hfill
	\subfloat[\label{1b}]{
		\includegraphics[scale=0.5]{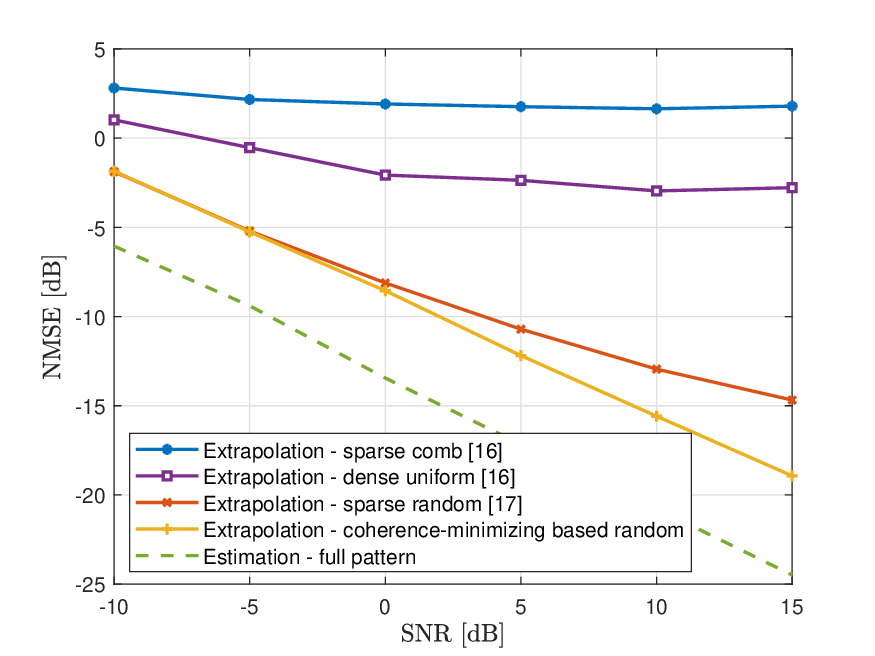} }
	\caption{\small{NMSE performance against the SNR for different antenna selection pattern designs, with (a) $\eta$ = 8, (b) $\eta$ = 4.}}
	\label{Figure_5} 
\end{figure}

Fig.~\ref{Figure_5} depicts the NMSE performance for both channel estimation and extrapolation under two settings (a) $\eta$ = 8, (b) $\eta$ = 4. These numerical results coincide well with the analytical results in Section IV-C. As expected, the random-position setups significantly outperform the deterministic-position setups. Among the deterministic setups, the dense uniform pattern exhibits better extrapolation performance than the sparse comb pattern, primarily due to its relatively lower side lobe levels. Although the sparse comb pattern exhibits an extremely narrow main lobe, it still suffers from $\eta$ strong side lobes, which severely degrade extrapolation performance. Without the coherence-minimizing procedure, the sparse random pattern exhibits instability and generally performs worse than the coherence-minimizing-based random pattern, a result attributed to the inherent randomness of this strategy. Specifically, at an SNR of 15 $\mathrm{dB}$, the performance gap between the sparse random pattern and the coherence-minimizing-based random pattern is 4.17 $\mathrm{dB}$ for $\eta$ = 8 and 4.25 $\mathrm{dB}$ for $\eta$ = 4. It is observed that the coherence-minimizing-based random pattern achieves the best extrapolation accuracy among all sparse patterns, although it remains inferior to channel estimation exploiting the full antenna array. This performance gain stems from its coherence-minimizing strategy, which generates a diverse set of candidate patterns and selects the one with the lowest mutual coherence, thereby enhancing angular and distance resolution and ensuring stable performance.
\color{black}
\begin{figure}[t]
    \centering
    \subfloat[\label{1a}]{
        \includegraphics[width=0.4\linewidth]{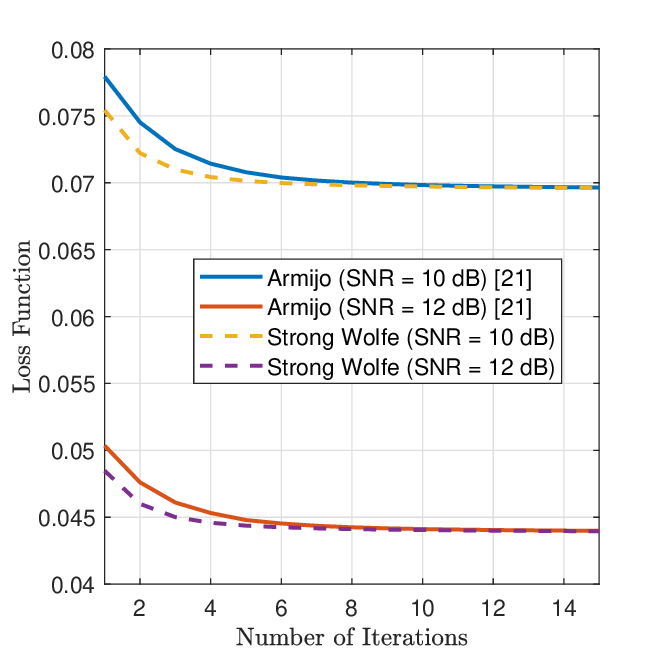}
    } \hfill
    \subfloat[\label{1b}]{
        \includegraphics[width=0.4\linewidth]{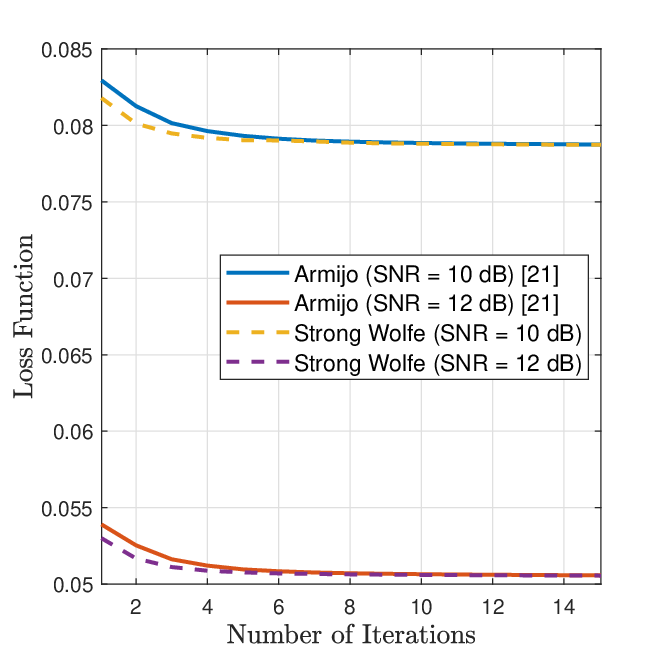}
    }
    \caption{\small{Convergence analysis of objective function using different line search methods, with (a) $\eta$ = 8, (b) $\eta$ = 4.}}
    \label{Figure_6}
\end{figure}

\par{\it{1) Convergence:}} The convergence behavior of the proposed P-ASIGW algorithm with different line search methods has been compared in Fig.~\ref{Figure_6} when (a) $\eta$ = 8, (b) $\eta$ = 4. The objective function $\|{\mathbf{H}}^\mathcal{B}-\tilde{{\boldsymbol{\Psi}}}(\hat{\boldsymbol{\theta}},\hat{\mathbf{r}})\hat{\mathbf{G}}\|_{F}^{2}$ exhibits a monotonic decrease over iterations. This indicates that the steps for updating $\hat{\boldsymbol{\theta}}$, $\hat{\mathbf{r}}$, and $\hat{\mathbf{G}}$ are monotonically non-increasing, ensuring that the alternating optimization procedure will converge, as supported by \cite{polar_dai}. Note that the line search method based on the strong Wolfe conditions achieves faster convergence than the Armijo condition-based method, as it incorporates an additional curvature condition, which facilitates the determination of an appropriate step size. Therefore, the strong Wolfe conditions-based method requires only 5 iterations on average, and the Armijo condition-based method necessitates 10 iterations on average to achieve comparable results.
\par{\it{2) Complexity:}} 
For the proposed P-ASOMP as stated in \textbf{Algorithm 1}, the computational complexity is dominated by operations of the SOMP procedure, i.e., steps 2-9 in \textbf{Algorithm 1}. Then we calculate the complexity of each step. Since ${\mathbf{\Psi}}\in\mathbb{C}^{K\times{N_aN_d}}$, ${\mathbf{R}}\in\mathbb{C}^{K\times{M}}$, and $\mathbf{H}^\mathcal{B}\in\mathbb{C}^{K\times{M}}$, the computational complexities from steps 3-8 are $\mathcal{O}(1)$, $\mathcal{O}(KN_aN_dM)$, $\mathcal{O}(N_aN_dM)$, $\mathcal{O}(1)$, $\mathcal{O}(\hat{L}^2KM+\hat{L}^3)$, and $\mathcal{O}(\hat{L}KM)$, respectively. Generally, the number of paths $\hat{L}$ is relatively small, thus the computation complexity from steps 3-8 is determined by $\mathcal{O}(KN_aN_dM)+\mathcal{O}(\hat{L}^2KM)$. After $\hat{L}$ iterations, the overall computational complexity is $\mathcal{O}(\hat{L}KN_aN_dM)+\mathcal{O}(\hat{L}^3KM)$. For the P-ASIGW algorithm in \textbf{Algorithm 2}, the total complexity includes the same initialization cost as P-ASOMP and a refinement stage with dominant operations in updating $\hat{\boldsymbol{\theta}}, \hat{\mathbf{r}}, \hat{\mathbf{G}}$. As shown in (13), the complexity of the update $\hat{\mathbf{G}}$ is $\mathcal{O}(\hat{L}^2KM+\hat{L}^3)$. Furthermore, the complexities of (15) and (16) are determined to be $\mathcal{O}(K^2M+KM^2+\hat{L}^2KM)$, benefiting from the small value of $\hat{L}$. After $N_\mathrm{iter}$ iterations, the complexity in the refinement stage becomes $\mathcal{O}((K^2M+KM^2+\hat{L}^2KM)N_\mathrm{iter})$. Consequently, the overall computational complexity of the proposed P-ASIGW method is $\mathcal{O}(\hat{L}KN_aN_dM)+\mathcal{O}(\hat{L}^3KM)+\mathcal{O}((K^2M+KM^2+\hat{L}^2KM)N_\mathrm{iter})$. Similarly, the proposed CV-P-ASOMP is stated in \textbf{Algorithm 3} exhibits the same structure as P-ASOMP, but restricts correlation operations to training subcarriers ($T \ll M$), reducing the complexity to $\mathcal{O}(\hat{L}KN_aN_dT + \hat{L}^3KM)$. Finally, the proposed CV-P-ASIGW in \textbf{Algorithm 4} combines the initialization cost from CV-P-ASOMP and a refinement process similar to P-ASIGW, resulting in an overall complexity of $\mathcal{O}(\hat{L}KN_aN_dT)+\mathcal{O}(\hat{L}^3KM)+\mathcal{O}((K^2M+KM^2+\hat{L}^2KM)N_\mathrm{iter})$.
\vspace{-0.5em}
\subsection{Evaluation of Channel Extrapolation Algorithms}
Having verified the effectiveness of coherence-minimizing-based random pattern design, we further compare our proposed algorithms with state-of-the-art OMP-based methods, such as P-OMP \cite{wideband_dai}, P-SOMP \cite{polar_dai}, and P-SIGW \cite{polar_dai}, exploiting this pattern design. The normalized mean square error (NMSE) of the extrapolated channel is calculated by ${\mathrm{NMSE}=\mathbb{E}(\frac{||{\mathbf{H}}^\mathcal{A}-\hat{\mathbf{H}}^\mathcal{A}||_{F}^{2}}{||{\mathbf{H}}^\mathcal{A}||_{F}^{2}})}$. 
\par Moreover, we test the system achievable rate using the extrapolated channel matrix. Let the extrapolated uplink channel matrix on the $m$-th subcarrier be denoted by $\hat{\mathbf{H}}_m = \left[ \hat{\mathbf{h}}_{m1}, \hat{\mathbf{h}}_{m2}, \dots, \hat{\mathbf{h}}_{mU} \right] \in \mathbb{C}^{N \times U}$, where $\hat{\mathbf{h}}_{mu}\in\mathbb{C}^{N \times 1}$ denotes the extrapolated channel vector between the $u$-th UE and the BS. The actual channel of the $x$-th user on the $m$-th subcarrier is represented by $\mathbf{h}_{mx}\in\mathbb{C}^{N \times 1}$, and the transmitted uplink signal from UE $u$ is $s_{mu} \sim \mathcal{CN}(0, p_t)$, where $p_t = \mathbb{E}[|s_{mu}|^2]$. Specifically, the zero-forcing (ZF) combining\footnote{To mitigate inter-user interference, ZF combining is employed at the receiver in this work.} vector for the $u$-th UE is then given by $\mathbf{v}_{mu} = \left[\hat{\mathbf{H}}_m \left( \hat{\mathbf{H}}_m^\mathrm{H} \hat{\mathbf{H}}_m \right)^{-1}\right](:, u)$. Based on this, the ergodic achievable uplink rate $R_u \in \mathbb{C}$ for the $u$-th UE can be formulated as (25), where the expectation is with respect to the channel estimates \cite{se_emil}, and $\tilde{\mathbf{h}}_{mu}\in\mathbb{C}^{N \times 1}$ denotes the channel extrapolation error, i.e., $\mathbf{h}_{mu} = \hat{\mathbf{h}}_{mu} + \tilde{\mathbf{h}}_{mu}$.
\begin{figure*}[!hb]
\normalsize
\setcounter{equation}{24} 
\begin{equation}
\label{*}
\begin{aligned}
R_u=\frac{1}{M}\sum_{m=1}^{M}\mathbb{E}\left\{\log_{2}\left(1+\frac{p_{mu}|\mathbf{v}_{mu}^\mathrm{H} \hat{\mathbf{h}}_{mu}|^{2}}{p_{mu} |\mathbf{v}_{mu}^\mathrm{H} \tilde{\mathbf{h}}_{mu}|^{2} + \sum_{x=1,x\neq{u}}^{U} p_{mx} |\mathbf{v}_{mu}^\mathrm{H} \mathbf{h}_{mx}|^{2} + \sigma^{2} \|\mathbf{v}_{mu}\|^{2}}\right)\right\}.
\end{aligned}
\end{equation}
\end{figure*}
\begin{figure*}[t!]
\begin{minipage}[t]{0.5\linewidth}
\centering
\includegraphics[width=1\linewidth]{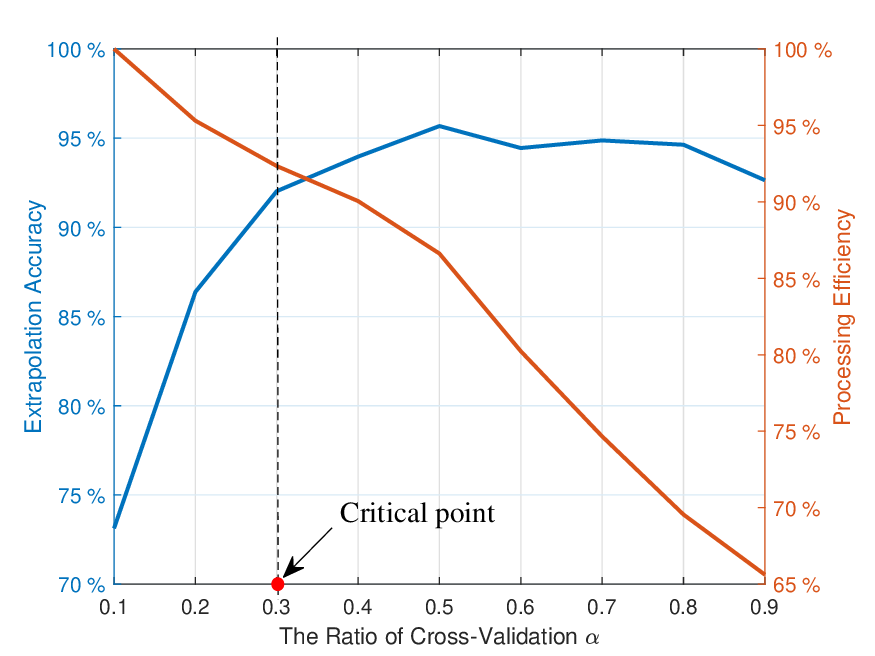}
\vspace{-1em}
\caption{\small{Comparison of extrapolation accuracy and processing efficiency against the CV ratio, with $\eta$ = 8.}\protect\label{Figure_7}}
\end{minipage}
\begin{minipage}[t]{0.5\linewidth}
\centering
\includegraphics[width=1\linewidth]{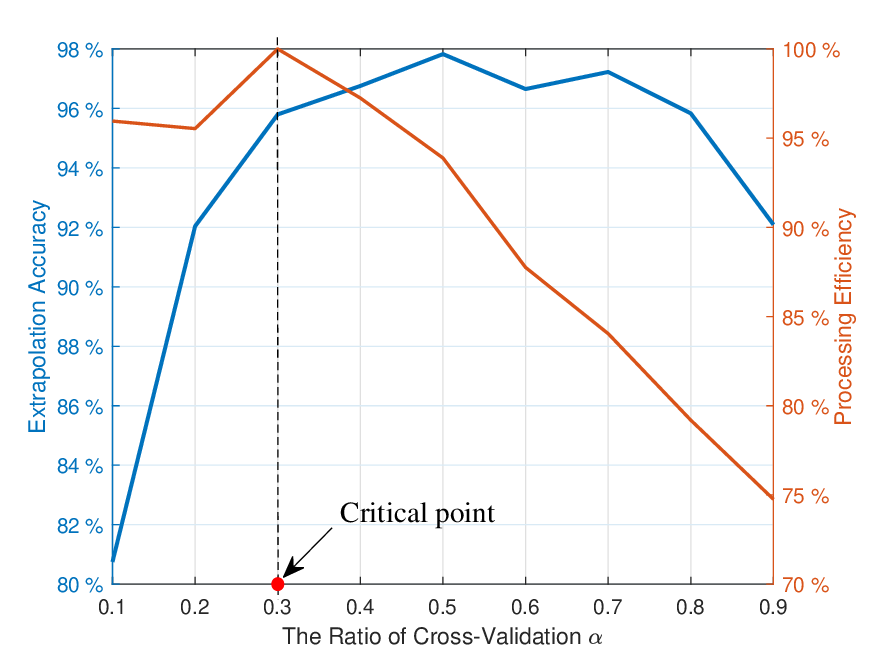}
\vspace{-1em}
\caption{\small{Comparison of extrapolation accuracy and processing efficiency against the CV ratio, with $\eta$ = 4.}\protect\label{Figure_8}}
\end{minipage}
\end{figure*}
\par Fig.~\ref{Figure_7} and Fig.~\ref{Figure_8} present the channel extrapolation accuracy and processing efficiency against the ratio of cross-validation $\alpha$ at the SNR of 0 $\mathrm{dB}$, for $\eta = 8$ and $\eta = 4$, respectively. The channel extrapolation accuracy, which quantifies the gap between P-ASOMP and CV-P-ASOMP, is calculated as $(1-\frac{|\mathrm{NMSE}\text{(P-ASOMP)}-\mathrm{NMSE}\text{(CV-P-ASOMP)}|}{\mathrm{NMSE}\text{(P-ASOMP)}})\times100\%$. The processing efficiency is computed as $\frac{t_{\mathrm{min}}}{t_{\alpha}} \times 100\%$, where $t_{\mathrm{min}}$ represents the minimal running time across different CV ratios, and $t_{\alpha}$ denotes the running time for the current CV ratio $\alpha$.
\color{black}
As illustrated in both Fig.~\ref{Figure_7} and Fig.~\ref{Figure_8}, $\alpha = 0.3$ serves as a critical point, achieving relatively high processing efficiency while maintaining an extrapolation accuracy exceeding ${90\%}$.
It is evident that larger values of $\alpha$ tend to reduce processing efficiency due to the increased computational cost associated with high-dimensional matrix operations. Conversely, excessively small values of $\alpha$ result in insufficient training samples, which may degrade the accuracy of channel extrapolation. Moreover, inadequate training data may cause an overestimation of the number of paths, thereby increasing the number of iterations and further exacerbating computational complexity, as shown in Fig.~\ref{Figure_8}.
\color{black}
Hence, to strike an effective balance between computational efficiency and extrapolation accuracy, the cross-validation ratio $\alpha$ is optimally chosen within the range $[0.3, 0.5]$. In the sequel, we adopt $\alpha = 0.3$ to ensure higher processing efficiency.
\begin{figure*}[t!]
\begin{minipage}[t]{0.5\linewidth}
\centering
\includegraphics[width=1\linewidth]{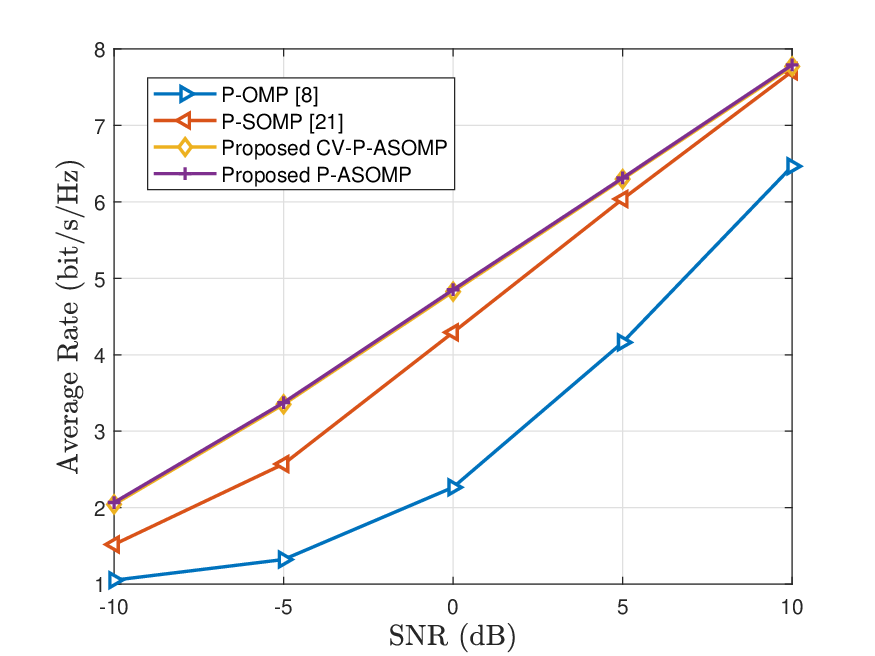}
\vspace{-1em}
\caption{\small{The average rates of various on-grid methods against the SNR, with $\eta$ = 4, $U$ = 3.}\protect\label{Figure_9}}
\end{minipage}
\begin{minipage}[t]{0.5\linewidth}
\centering
\includegraphics[width=1\linewidth]{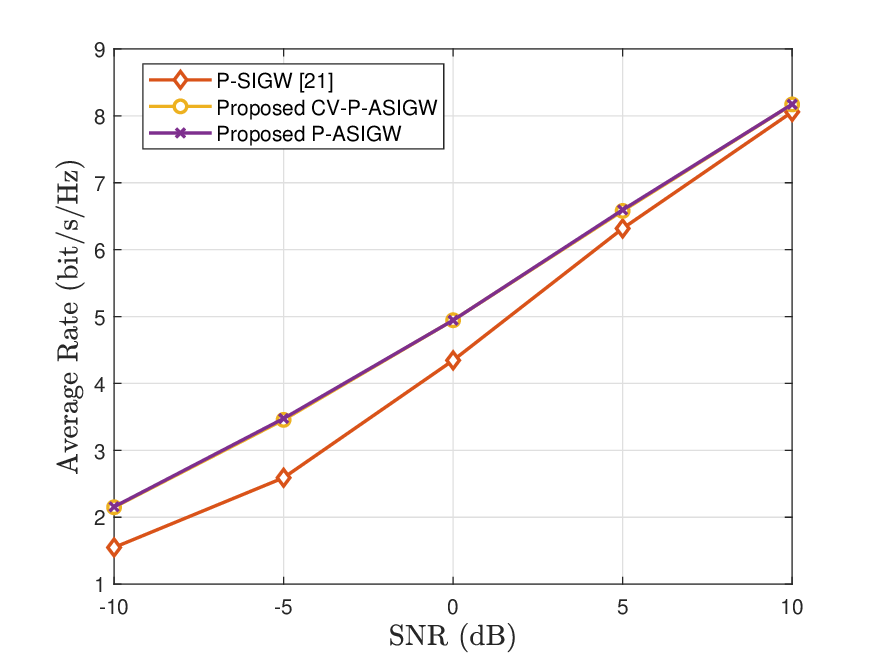}
\vspace{-1em}
\caption{\small{The average rates of various off-grid methods against the SNR, with $\eta$ = 4, $U$ = 3.}\protect\label{Figure_10}}
\end{minipage}
\end{figure*}
\par The average uplink achievable rates of various on-grid methods are plotted in Fig.~\ref{Figure_9}, while those of off-grid methods are shown in Fig.~\ref{Figure_10}, with $U = 3$ and $\eta = 4$, respectively. 
The first observation is that the proposed P-ASOMP and CV-P-ASOMP algorithms outperform existing on-grid methods in terms of achievable rates across all SNR regimes, demonstrating that the adaptive channel extrapolation scheme can effectively exploit the inherent channel sparsity. Moreover, the proposed P-ASOMP and CV-P-ASOMP algorithms exhibit comparable extrapolation performance to the P-SOMP \cite{polar_dai} algorithm under high SNR conditions. This is because the iteration number in the P-SOMP \cite{polar_dai} method is specifically set for high SNR scenarios. Based on Fig.~\ref{Figure_9}, it can be observed that the on-grid P-ASOMP and CV-P-ASOMP algorithms exhibit nearly identical performance in terms of average rates. Similarly, as shown in Fig.~\ref{Figure_10}, the off-grid P-ASIGW and CV-P-ASIGW algorithms also demonstrate almost the same performance. This result highlights the effectiveness of the proposed CV-based scheme in significantly reducing computational complexity while preserving high extrapolation accuracy.
\begin{figure}[t]
\centering
\includegraphics[width=0.55\linewidth]{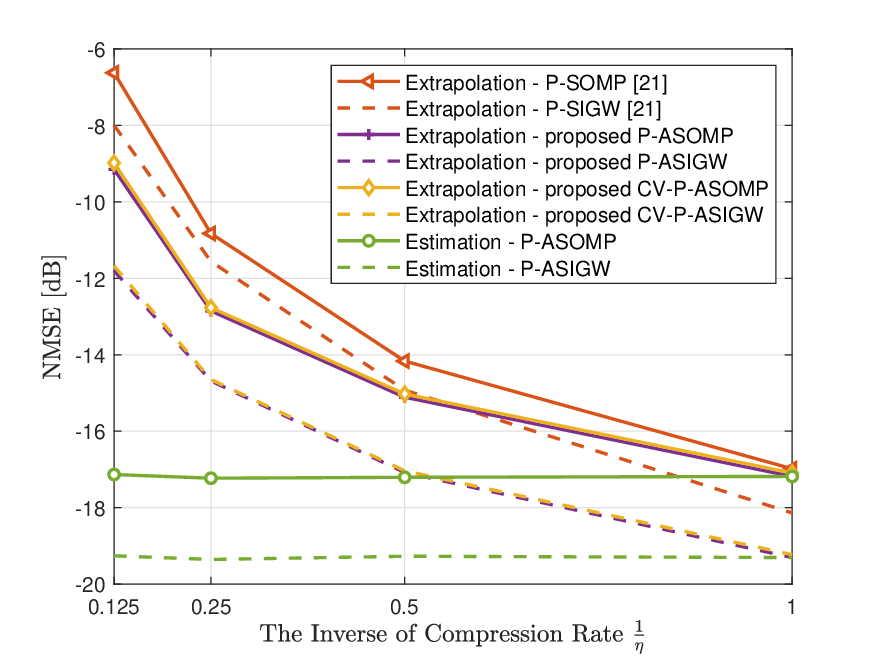}
\vspace{-1em}
\caption{\centering{\small{NMSE values of different channel extrapolation algorithms across various compression rates $\eta$.} \protect\label{Figure_11}}}
\end{figure}
\par In Fig.~\ref{Figure_11}, the NMSE performance of the on-grid and off-grid methods is plotted against the compression ratio $\eta$ at a SNR of 5 $\mathrm{dB}$. As illustrated in Fig.~\ref{Figure_6}, the line search method based on the strong Wolfe conditions exhibits faster convergence compared to its counterpart relying on the Armijo condition \cite{polar_dai}. Given that the P-SIGW \cite{polar_dai} algorithm employs an Armijo condition-based line search, it requires a higher number of iterations for convergence. In contrast, the proposed P-ASIGW and CV-P-ASIGW algorithms incorporate a strong Wolfe conditions-based line search strategy, which significantly reduces the required number of iterations for convergence. In this paper, the maximum iteration count is set to 10 for the P-SIGW \cite{polar_dai} algorithm and reduced to 5 for the proposed P-ASIGW and CV-P-ASIGW algorithms. By refining angle and distance parameters with significantly higher resolution, the off-grid algorithms achieve substantially improved NMSE performance compared to the on-grid algorithms. Specifically, when the inverse of the compression rate is set to $0.5$, the performance gap between P-SOMP \cite{polar_dai} and P-SIGW \cite{polar_dai} is approximately 0.76 dB. In contrast, the performance gaps between P-ASOMP and P-ASIGW, as well as between CV-P-ASOMP and CV-P-ASIGW, both reach 2.01 dB. Furthermore, it is observed that the proposed adaptive off-grid algorithms, namely P-ASIGW and CV-P-ASIGW, clearly outperform the fixed-iteration off-grid algorithm P-SIGW \cite{polar_dai}. This improvement is primarily attributed to the adaptive mechanism, which enables more accurate initial value estimation, thereby enhancing the overall reconstruction accuracy.
\begin{figure}[t]
\centering
\begin{minipage}[t]{0.48\linewidth}
    \centering
    \includegraphics[width=\linewidth]{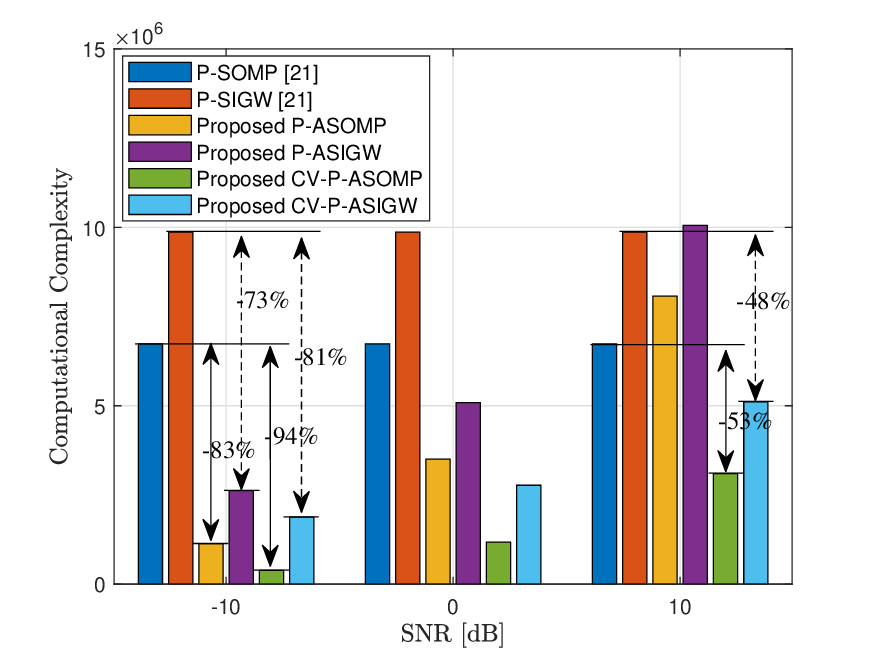}
    \caption{Computational complexity of various channel extrapolation algorithms, with $\eta = 8$.}
    \label{Figure_12}
\end{minipage}
\hfill
\begin{minipage}[t]{0.48\linewidth}
    \centering
    \includegraphics[width=\linewidth]{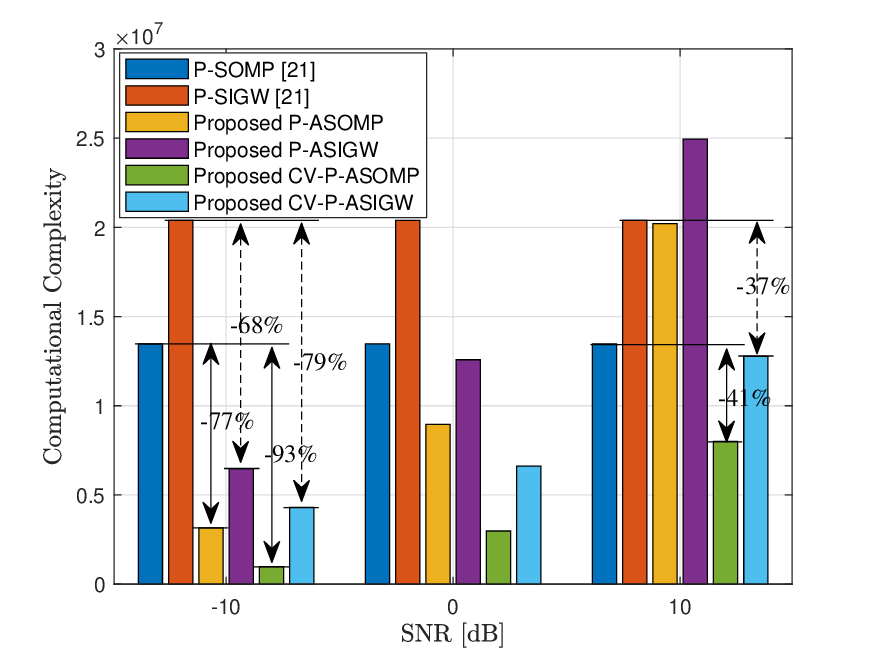}
    \caption{Computational complexity of various channel extrapolation algorithms, with $\eta = 4$.}
    \label{Figure_13}
\end{minipage}
\end{figure}

\par Fig.~\ref{Figure_12} and Fig.~\ref{Figure_13} illustrate the computational complexity of various channel extrapolation algorithms, with $\eta$ = 8 for Fig.~\ref{Figure_12}, and $\eta$ = 4 for Fig.~\ref{Figure_13}. Since the optimal number of iterations varies across the SNR range, the computational complexity of the proposed P-ASOMP, P-ASIGW, CV-P-ASOMP, and CV-P-ASIGW algorithms increases with rising SNR. In low SNR scenarios, the proposed on-grid and off-grid algorithms exhibit superior performance by employing a well-designed stopping criterion, which minimizes the need for extensive iterations. As shown in Fig.~\ref{Figure_12}, at an SNR of -10 $\mathrm{dB}$, the P-ASOMP and CV-P-ASOMP algorithms reduce computational complexity by approximately 83\% and 94\% compared to the P-SOMP \cite{polar_dai} method, respectively. For the off-grid methods, the P-ASIGW and CV-P-ASIGW algorithms achieve reductions of around 73\% and 81\% compared to the P-SIGW \cite{polar_dai} method, respectively. As depicted in Fig.~\ref{Figure_11} and Fig.~\ref{Figure_12}, the proposed adaptive scheme effectively captures the sparsity condition, while the CV-based adaptive scheme achieves satisfactory extrapolation performance with low computational complexity. At a SNR of 10 $\mathrm{dB}$, the computational complexity of the proposed P-ASOMP and P-ASIGW algorithms increases significantly due to the rise in iteration numbers. However, the CV-based scheme effectively alleviates this issue, achieving comparable extrapolation performance while reducing computational complexity by approximately 50\%. Additionally, computational complexity increases as the number of active elements grows, which is attributed to higher-dimensional matrix computations. In this case, the on-grid and off-grid methods based on the CV scheme still demonstrate significant advantages. As shown in Fig.~\ref{Figure_13}, even in the 10 $\mathrm{dB}$ scenarios with a high number of iterations, the CV-P-ASOMP and CV-P-ASIGW algorithms reduce computational complexity by about 41\% and 37\% compared to the P-SOMP \cite{polar_dai} and P-SIGW \cite{polar_dai} methods, respectively. Notably, by selecting an appropriate CV ratio, the CV-P-ASOMP algorithm achieves performance comparable to the P-ASOMP algorithm while significantly improving processing efficiency. This is attributed to the shared support sets across all subcarriers, enabling the CV-based algorithms to reduce high-dimensional matrix computations in the correlation step by leveraging shared information in the polar-domain.
\color{black}

\section{Conclusion}\label{se:model}
In this paper, we formulated the channel extrapolation problem as a sparse signal recovery task and proposed a compressed sensing-based spatial-domain channel extrapolation framework for mmWave dynamic XL-MIMO communication systems. Then, we developed both on-grid algorithms, namely P-ASOMP and CV-P-ASOMP, and off-grid algorithms, namely P-ASIGW and CV-P-ASIGW, for effective near-field channel extrapolation with low computational complexity. Under this framework, the measurement matrix was constructed based on a strategically selected subset of antenna elements. By leveraging a pre-designed polar-domain dictionary to accurately characterize the inherent spatial correlation of near-field channels, we observed that the mutual coherence of the sensing matrix is primarily determined by the antenna selection pattern. Since lower mutual coherence directly implies improved reconstruction performance, we introduced a coherence-minimizing-based random pattern design to enhance extrapolation accuracy. Building on this pattern design, we evaluated the proposed on-grid and off-grid algorithms. In particular, the off-grid P-ASIGW and CV-P-ASIGW methods incorporate a strong Wolfe conditions-based line search to accelerate convergence. Notably, the proposed CV-based scheme can maintain satisfactory channel extrapolation accuracy under the MMV framework, while significantly reducing computational complexity compared to existing methods.
\color{black}

\bibliographystyle{ieeetr}
\bibliography{bibliography.bib}

\end{document}